\documentclass[aps,prb,twocolumn,showpacs,amsmath,amssymb,floats,showkeys,preprintnumbers,superscriptaddress]{revtex4-1}
\usepackage[dvips]{graphicx}
\usepackage{color}
\usepackage{float}
\usepackage{epsfig}
\usepackage{epstopdf}
\usepackage{psfrag}
\usepackage{ulem}
\usepackage{multirow}
\usepackage{dcolumn}
\usepackage{hyperref}
\usepackage[usenames,dvipsnames,svgnames,table]{xcolor}

\newcommand{\be}{\begin{equation}}
\newcommand{\ee}{\end{equation}}
\newcommand{\bea}{\begin{eqnarray}}
\newcommand{\eea}{\end{eqnarray}}
\newcommand{\ba}{\begin{eqnarray*}}
\newcommand{\ea}{\end{eqnarray*}}
\newcommand{\dagga}{{\phantom{\dagger}}}
\newcommand{\bR}{\mathbf{R}}

\newcommand{\bq}{\mathbf{q}}

\newcommand{\bk}{\mathbf{k}}

\newcommand{\up}{\uparrow}
\newcommand{\down}{\downarrow}

\newcommand{\eqn}[1]{(\ref{#1})}

\begin{document}
 
\title{Electron doped organics: charge-disproportionate insulators and Hubbard-Fr{\"o}hlich metals}

\author{S. Shahab Naghavi}
\affiliation{International School for Advanced Studies (SISSA), and CNR-IOM Democritos National Simulation Center, Via Bonomea 265, I-34136 Trieste, Italy }
\author{Michele Fabrizio}
\affiliation{International School for Advanced Studies (SISSA), and CNR-IOM Democritos National Simulation Center, Via Bonomea 265, I-34136 Trieste, Italy }
\author{Tao Qin}
\affiliation{International School for Advanced Studies (SISSA), and CNR-IOM Democritos National Simulation Center, Via Bonomea 265, I-34136 Trieste, Italy }
\author{Erio Tosatti}
\affiliation{International School for Advanced Studies (SISSA), and CNR-IOM Democritos National Simulation Center, Via Bonomea 265, I-34136 Trieste, Italy }
\affiliation{International Centre for Theoretical Physics (ICTP), Strada Costiera 11, I-34151 Trieste, Italy 
}
\date{\today}
\begin{abstract} 
  Several  examples of  metallic  electron  doped polycyclic  aromatic
hydrocarbons   (PAHs)   molecular    crystals   have   recently   been
experimentally   proposed.    Some   of  them   have   superconducting
components, but  most other details  are still unknown  beginning with
structure  and  the  nature  of   metallicity.  We  carried  out  {\sl
ab-initio}   density  functional   calculations  for   La-Phenanthrene
(La-PA),  here meant  to represent  a generic  case of  three-electron
doping,  to investigate  structure  and properties  of a  conceptually
simple case.   To our surprise we  found first of all  that the lowest
energy   state  is   not  metallic   but  band   insulating,  with   a
disproportionation of  two inequivalent  PA molecular  ions and  a low
$P1$ symmetry, questioning the  common assumption that three electrons
will  automatically  metallize  a  PAH  crystal.   Our  best  metallic
structure is  metastable and  slightly higher  in energy,  and retains
equivalent  PA  ions  and  a  higher $P2_{1}$  symmetry  --  the  same
generally claimed  for metallic  PAHs.  We  show that  a ``dimerizing"
periodic  distortion  opens very  effectively  a  gap  in place  of  a
symmetry related degeneracy of all  $P2_{1}$ structures near the Fermi
level, foreshadowing  a possible  role of that  special intermolecular
phonon in superconductivity of  metallic PAHs.  A Hubbard-Fr{\"o}hlich
model describing that situation is formulated for future studies.
\end{abstract}

\maketitle

\section{introduction}
\label{SEC:INTRO}
There is a continuing and expanding interest in the transformation, via electron
doping, of insulating organic molecular crystals to molecular metals and especially 
to superconductors.  Polycyclic aromatic hydrocarbons (PAHs) are a well known 
family of molecular crystals where intercalation of electron-donating atomic species 
recently rekindled attention in that direction. Pioneering reports include 
metallization with superconductivity in electron doped picene,\cite{mitsuhashi10} 
coronene,\cite{kubozono11, kato11} dibenzopentacene, where $T_{\rm c}$ exceeds 
30\,K,\cite{xue12} and phenanthrene.\cite{wang11, wang11a, wang12} To this date, many 
of these exciting systems are proving difficult to reproduce and characterize, and 
there remains considerable uncertainty across the board, ranging from precise 
stoichiometry, chemical composition, to crystal and electronic structure, the role of 
correlations, to superconductivity and its mechanism. Surface spectroscopies, 
intriguingly, generally fail even to show metallicity of the doped materials.\cite{mahns12,caputo12}
Early calculations\cite{subedi11, giovannetti11,casula11,kato11,kosugi11,Andres11,Picene_abinitio} 
illustrated partial filling of LUMO+1 derived bands, weakly hybridized with higher energy alkali states.
Electron-phonon calculations highlighted the coupling of these bands to
intra-molecular but also inter-molecular lattice vibrations, potentially conducive 
to BCS superconductivity.\cite{casula11} At the same time, large electron-electron 
repulsion parameters have been repeatedly 
emphasized,\cite{giovannetti11, kosugi11, huang12} suggesting possible analogies with 
alkali fullerides--molecular superconducting systems\cite{hebard91} earlier proposed\cite{capone02} and recently 
shown\cite{ganin08,ganin10} to be strongly correlated superconductors, close 
to a Mott-insulating state.\cite{caponeRPM09} All that leaves the overall 
nature of doped PAHs, their insulating, metallic and superconducting phases, in a state 
of lamentable uncertainty. 

Here we attempt a fresh theoretical start, based on accurate {\sl ab-initio} total energy 
calculations and structural optimization of a particularly simple doped crystalline PAH system. 
As an alternative to previous studies, which also included a careful optimization of K-doped 
picene for various potassium concentrations,\cite{kosugi11} we choose here the
more speculative case of La-Phenanthrene (La-PA).  With the single electropositive trivalent
atom and the more compact PA molecule accepting or sharing three electrons, this is 
a conceptually and practically much simpler system.  La-PA is reported to exist, and 
to superconduct near 6\,K,\cite{wang12}
although this data has not been independently confirmed so far. In the context of a model theoretical study,
hypothetical crystalline La-PA serves well as an idealized model system, sharing many of the 
properties of the wider class of electron doped PAHs, with reduced complexity.  The PA molecule C$_{14}$H$_{10}$ 
is small and relatively rigid, and there is only one La cation to be located in the cell, next to it. 
This simplicity affords a much more intimate and exhaustive structural and electronic search.
In turn, the results and modeling can be expected to fulfill a broader explorative
scope, defining a prototype Hamiltonian of more general value, beyond the specific system chosen to define it. 

In section~\ref{SEC:DFT} we 
describe density functional theory  (DFT) calculations 
of total energy and electronic structure of La-PA as a function of all atomic 
coordinates in a bimolecular unit cell, with the scope of determining the nature of 
the lowest energy crystal structures and the corresponding  electronic energy bands. 
Rather unexpectedly, we find that the optimal crystal structure is insulating due to 
a spontaneous disproportionation, where La atom pairs preferentially bind to one PA molecule 
and less to the other. The spontaneous nonequivalence of the two PA molecules in the cell leads to a lower $P1$ lattice 
symmetry, directly leading to the opening of an insulating band gap. No such disproportionated state 
has so far emerged in previous theoretical studies of alkali doped PAHs, and in experimental studies; 
our results suggest that this possibility should be pursued.  
In our search for La-PA crystalline structures we also find at 
higher energy metallic structures with $P2_{1}$ symmetry, where the La 
cation is symmetrically equidistant among two PA anions. This high symmetry, similar to that proposed 
for many other postulated PAH superconductors, implies a symmetry-induced degeneracy near the Fermi level.  
We thus take it up this metallic phase and examine it as a simple prototype system, useful  for the future 
development of further generic theoretical 
modeling of metallic electron doped PAHs, independently of whether a phase with these or nearby characteristics 
will actually be confirmed or not in La-PA. In section~\ref{SEC:TBM}, we downfold the electronic structure 
into a tight binding minimal two-band model on a basis of localized Wannier functions.  
In section~\ref{SEC:DIMER}, we briefly discuss how a frozen intermolecular phonon 
consisting of a simple dimerization of PA molecular pairs constitutes a 
gap-opening mode for electrons near Fermi, within the two LUMO+1 derived and partly 
degenerate bands. Reminiscent of the situation of MgB$_2$, the linear band 
splitting caused by a specific periodic distortion, foreshadows 
a strong role of a particular intermolecular phonon with PA-dimerizing eigenvector, 
in a hypothetical BCS type superconducting state. The third ingredient, already highlighted by previous studies,
is the strong expected electron-electron repulsion within each molecular ion.  As our final result we assemble the tight binding 
model and the electron phonon together with the intramolecular Coulomb interaction parameter $U$, thus  forming 
what we designate a two-band {\it Hubbard-Fr{\"o}hlich model}, which we consider a minimal model for future studies of this kind of 
potential superconductor. Finally, in section~\ref{SEC:CONC} we draw our conclusions.

\section{DFT total energy calculations and structural optimization of La-PA}
\label{SEC:DFT}

\textit{Ab-initio} electronic structure calculations for La-PA were carried out 
using the {\sc Quantum-ESPRESSO}\cite{pwscf} code which implements the standard 
DFT framework within a plane waves basis set for one-electron wavefunctions. 
The Generalized Gradient Approximation (GGA) used a PBE exchange-correlation 
potential.\cite{PBE} La was treated with Vanderbilt ultrasoft pseudopotentials,\cite{PP-VDB} 
whereas the potentials for carbon and oxygen, C.pbe-rrjkus.UPF and H.pbe-rrjkus.UPF, were 
taken from the {\sc Quantum-ESPRESSO} web package. The $k$-point sampling for the 
integration in the Brillouin zone (BZ) yielded well converged results with 3$\times$4$\times$3  
Monkhorst-Pack $k$-point meshes. Plane wave cut-offs were 50\,eV  for kinetic 
energy and 500\,eV for charge density.  

\begin{table}[htb!]
\centering
\caption{Theoretically DFT optimized structural parameters of the pristine 
         phenanthrene molecular crystal, with $P2_{1}$ symmetry and two molecules 
         per cell, obtained with different approximations, compared with experiment. 
         Lengths in Angstrom units.}
\begin{ruledtabular}
\begin{tabular}{cccccccc}
Source  &  $a$  & $b$  & $c$ & $\alpha$ & $\beta$  &  $\gamma$ & Vol. \\ \hline
Exp.\cite{phen-crystal_siss}   & 8.46  &  6.16  &  9.47  &  90.0  &  97.7  &  90.0  &  489.1 \\
GGA      &   9.25  &  6.31  &  9.71  &  90.0  & 100.6  &  90.0  &  557.4         \\
vdW-DF\cite{vdW-DF_SISS}   &   8.51  &  6.19  &  9.49  &  90.0  &  98.2  &  90.0  &  494.8            \\
vdW-DF2\cite{vdW-DF2}  &   8.27  &  6.09  &  9.34  &  90.0  &  97.7  &  90.0  &  466.5           \\
Grimme\cite{vdW-parameters,vdW-Grimme_SISS}   &   7.87  &  5.95  &  9.19 &  90.0  &  96.5  &  90.0  &  427.7          \\
\end{tabular}
\end{ruledtabular}
\label{TAB:vdW}
\end{table}

As a preliminary step we began with pristine phenanthrene (PA). As for all 
molecular crystals, we face the problem of van der Waals (vdW) 
forces, which dominate large distance intermolecular interactions but are missing 
in standard DFT, with uncontrolled errors in the optimal geometry and in the 
electronic structure. Aiming at a quantitative description of the pristine PA 
molecular crystal structure we tried different vdW schemes among those currently proposed. 
Assuming a bimolecular $P2_{1}$ symmetry unit cell as known experimentally, we 
found best agreement with the experimental  structure parameters \cite{phen-crystal_siss} 
of  table \ref{TAB:vdW} by complementing DFT with the vdW-DF 
functional.\cite{vdW-DF_SISS,vdW-DF_SIS,vdW-DF_SI,vdW-Stress-pwscf} All energy 
calculations in the rest of this paper were carried out including the vdW-DF contribution.

\begin{figure}[htb!]
\centering
\includegraphics[width=.8\linewidth]{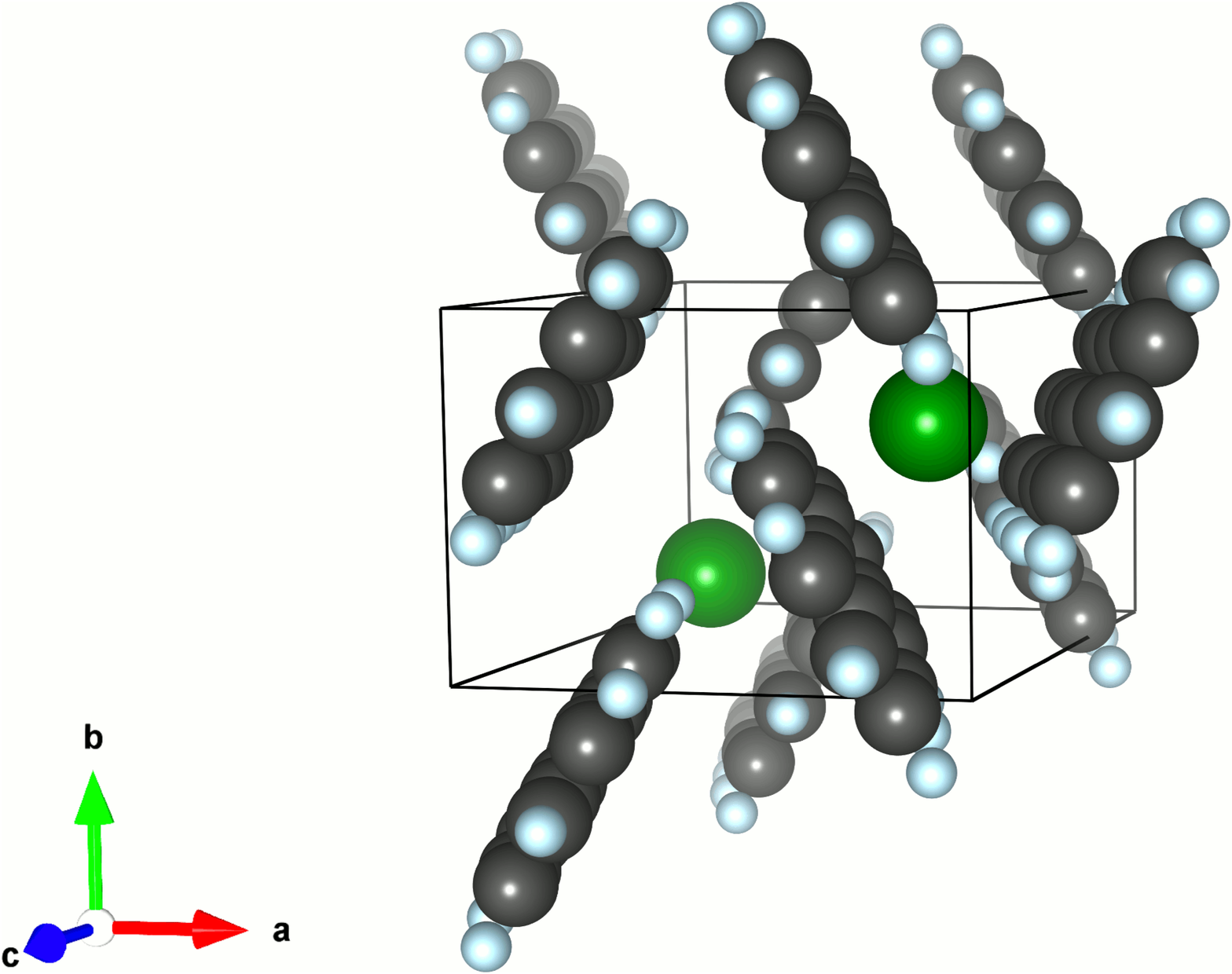}
\caption{(Color online) Schematic of one of 57  trial structures, with two
PA molecules in the unit-cell, two La atoms (green) and $P2_{1}$ symmetry.
These structures are subsequently relaxed, retaining at first a $P2_{1}$ symmetry.
From the few best ones, additional trial structures of lower $P1$ symmetry are generated
by exchanging La positions between them.}
\label{FIG:FILL}
\end{figure}

\begin{figure}[htb!]
\centering
\includegraphics[width=1.\linewidth]{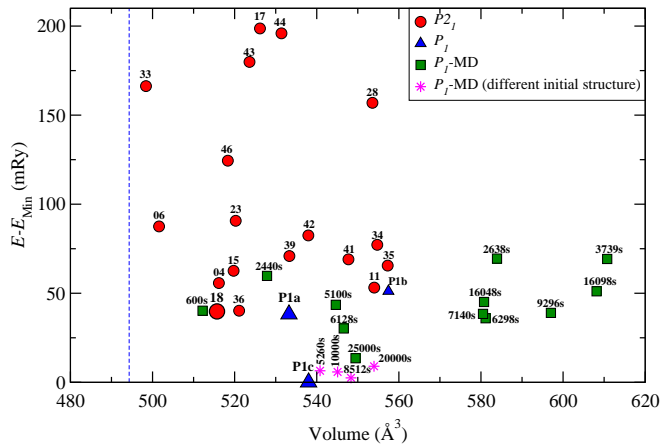}
\caption{(Color online) Energies and volumes per (bimolecular) cell for the main 
         relaxed structures. The  lowest energy structures of $P1$ symmetry are insulating, the
         metastable $P2_{1}$ structures are metallic. Magenta asterisks
         are MD-simulations starting from ``P1a" structure rather than 18, shown by blue triangle.
         The dashed line shows the 
         cell volume quoted in the experimental Ref. \onlinecite{wang12}.}
\label{FIG:ENERGYLAND}
\end{figure}

Thus equipped, we moved on to our main target, the electron-doped  La-PA system. The experimental suggestion 
\cite{wang12} that La-PA retains the same $P2_{1}$  symmetry of pristine PA was not forced; 
but we chose to restrict to the same minimal bimolecular unit cell.
To explore within that bound the largest variety of structural configurations, we initially placed two PA molecules 
in the pristine crystal positions. A large number $n$=114 of empty spaces that could host a La atom 
between the PA molecules were identified using  76 and 31\,pm respectively  for the covalent radii 
of carbon and hydrogen, and 100--117\,pm for the ionic radius of La$^{+3}$.\footnote{We thank A. Laio for this suggestion} 
If two of these empty sites per cell are to be filled with La atoms, that can in principle be done in 
$\frac{n(n-1)}{2}$= 6441 different ways. 
Even though several configurations such as those with two La atoms closest to each other 
could be discarded as very unlikely, the overall number of La placements is still too large to be explored in full. 
We therefore adopted a shortcut strategy. First we started, as suggested by the experimental 
papers, with $P2_{1}$ symmetry of the two-molecule unit cell, see e.g. Fig. \ref{FIG:FILL}, thus spanning a much smaller set 
of structures, which we can explore in full.  After that, random structures of lower ($P1$) 
symmetry were created by intermixing the position of La atoms taken from the few optimal 
high $P2_{1}$ symmetry structures. While of course this procedure
is not equivalent to 
a full search, it does quite well as we shall see, bringing out unmistakeable novelties. 
In the restricted $P2_{1}$ search, the first La is placed in $n$ possible ways, and the 
second La position is completely determined by symmetry, reducing the number of starting geometries 
from $\frac{n(n-1)}{2}$ to $n$=114. The $C_{2}$ rotation axis moreover reduces that number 
by a further factor 2, down to $n/2$ = 57 different initial $P2_{1}$ trial structures. For 
each of them we calculated the DFT electronic structure, total energy, and forces acting on 
all atoms. The forces were then used for a relaxation of all atomic positions. 
A full optimization---lattice parameters as well as all internal coordinates---was performed  
for all structures. At convergence, residual forces acting on each atom were 
much smaller than 1\,mRy/a.u..
During the relaxation process, the La atoms sometimes moved by large amounts, so that  
different initial configurations often ended up to the same final, optimized structure.
Therefore, at the end of optimization, all the 57 different structures 
were categorized to 18 different groups.  The parameters for a few the lowest energy structures 
are given in table~\ref{TAB:LANDSCAPE} and all are plotted in Fig.~\ref{FIG:ENERGYLAND}. 

\begin{table*}[htb!]
\centering
\caption{Energy and structural information of  $P2_{1}$ structures and $P1$ structures of lowes energy. 
Lengths are in  Angstrom and energies in mRy/bimolecular cell.}
\begin{ruledtabular}
    \begin{tabular}{lcccccccc}
    Samples &  $E-E_{18}$[mRy]    & $a$     & $b$  & $c$  & $\alpha$ & $\beta$  &  $\gamma$ & Volume \\ \hline
Exp.~\cite{wang12} & ......... & 8.481 & 6.187 &  9.512 & 90.00 & 97.95 & 90.00 & 494.3  \\ \hline
\multicolumn{9}{|c|}{$P2_{1} symmetry$} \\ \hline   
18     &    0.0  &    8.55  &   6.61  &  10.07  &  90.00  &  115.11   &  90.00  &  515.70   \\   
36     &    0.4  &    8.62  &   6.59  &   9.93  &  90.00  &  112.46   &  90.00  &  521.10   \\   
11     &   13.4  &    8.54  &   7.24  &   9.01  &  90.00  &   96.01   &  90.00  &  554.01   \\   
4      &   15.9  &    8.37  &   6.63  &   9.49  &  90.00  &  101.14   &  90.00  &  516.17   \\   \hline
\multicolumn{9}{|c|}{$P_{1} symmetry$} \\ \hline   
P1c    &  -39.8  &    9.73  &   5.83  &   9.73  &  86.12  &  101.89   &  93.26  &  538.01   \\   
8512s  &  -37.3  &    9.68  &   5.87  &  10.03  &  94.65  &  105.14   &  86.56  &  548.34   \\   
10000s &  -34.0  &    9.75  &   5.80  &  10.56  &  91.78  &  113.92   &  86.41  &  545.08   \\   
5260s  &  -33.3  &    9.36  &   5.99  &  10.09  &  87.99  &  106.21   &  86.44  &  540.81   \\ 
20000s &  -30.9  &    9.80  &   5.87  &  10.57  &  91.95  &  114.05   &  86.64  &  553.95   \\   
25000s$^{\rm a}$  &   -26.3  &     9.08 &   6.95  &  10.99  &  92.93  &  127.21   &  83.52  &  549.44   \\ 
    \end{tabular}
\footnotetext[1]{Initial structure of this MD simulation is 18, while for others is P1a}
\end{ruledtabular}
\label{TAB:LANDSCAPE}
\end{table*}

\begin{figure}[htp!]
\centering
\includegraphics[width=.66\linewidth]{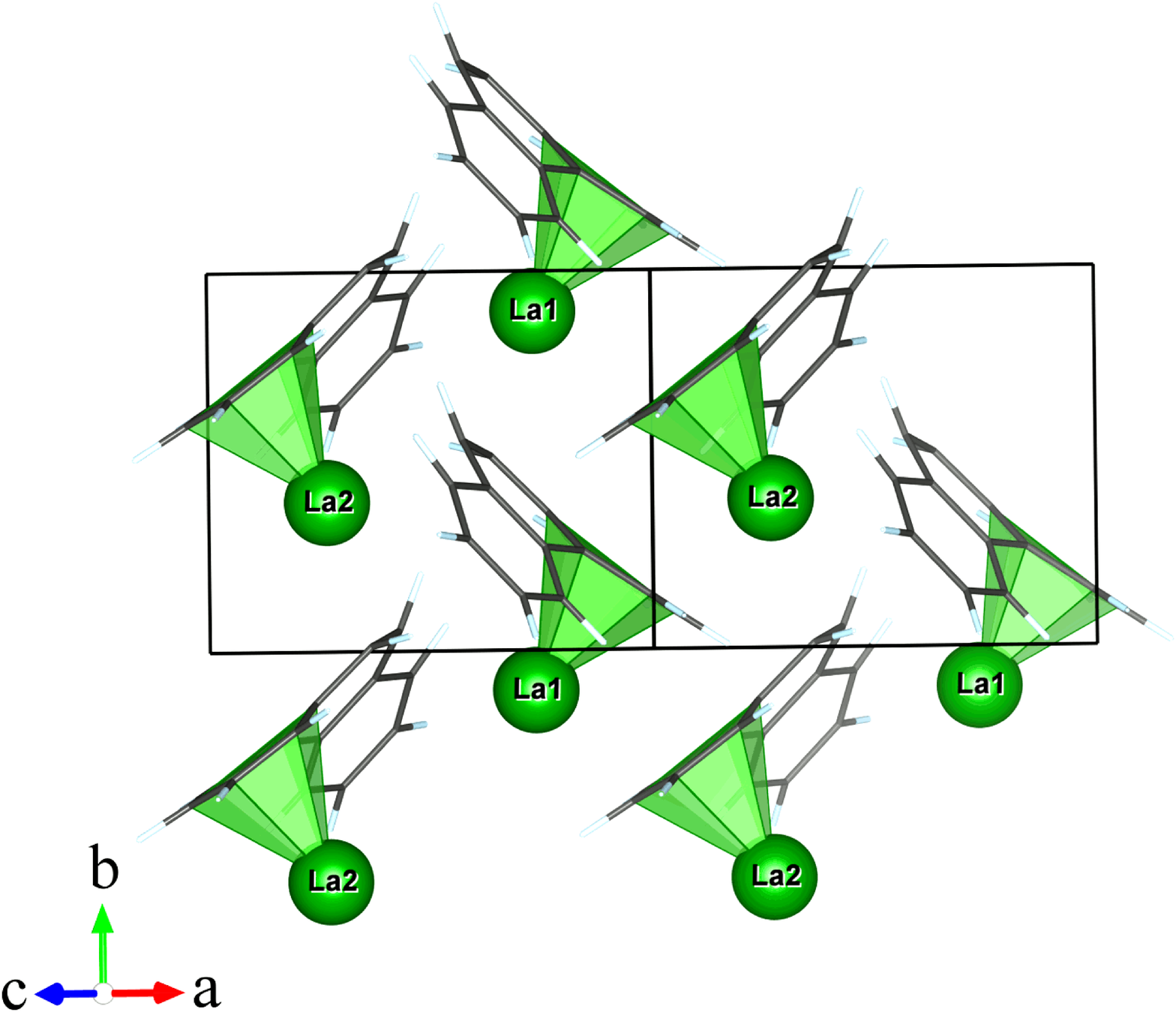}
\begin{center}
(a)
\end{center}
\includegraphics[width=.66\linewidth]{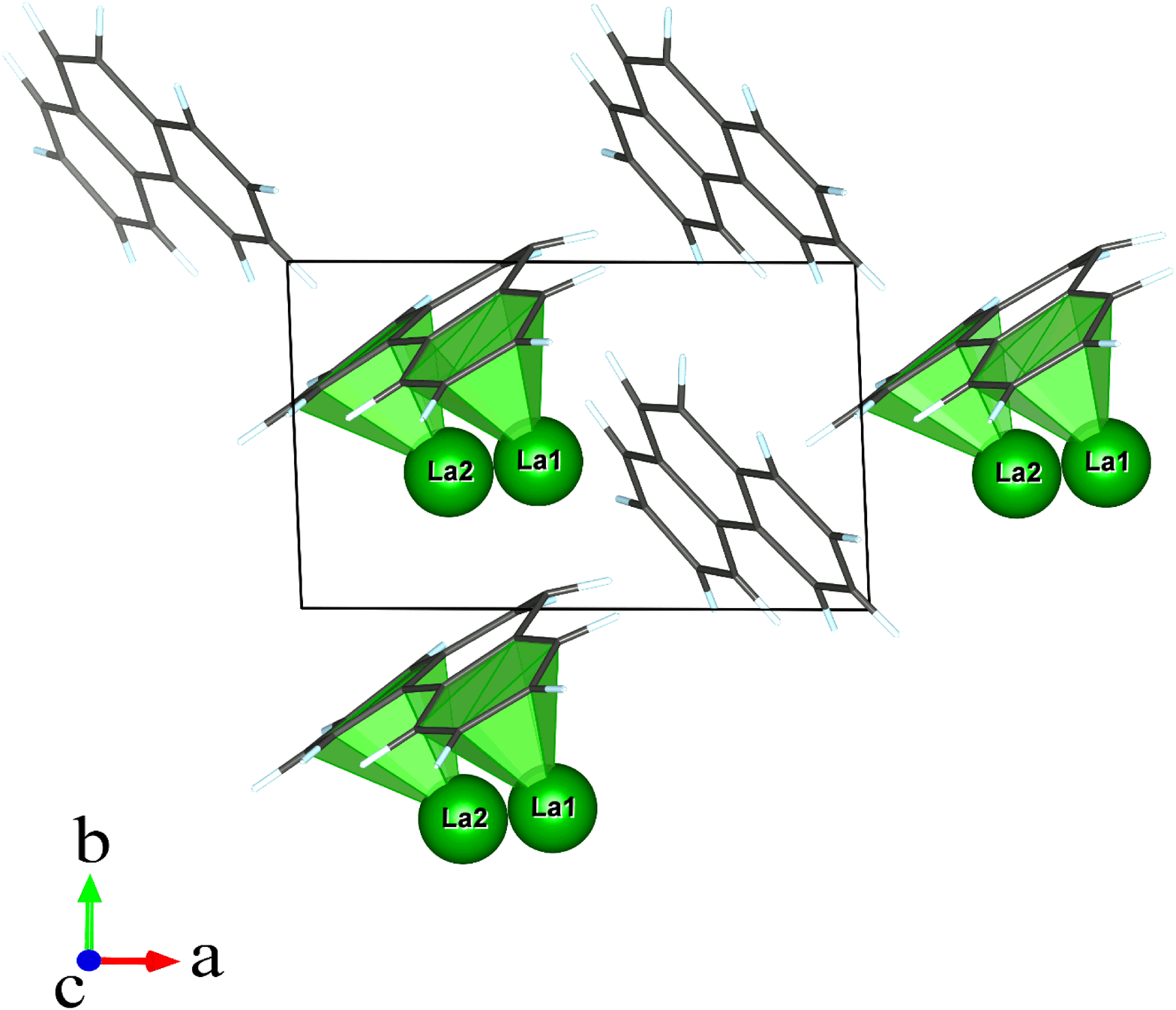}
\begin{center}
(b)
\end{center}
\caption{(Color online) (a): Schematic of the best metallic structure 18 ($P2_{1}$ symmetry, metastable).
         The two PA molecules are equivalent, and heavily distorted, with La atoms positioned symmetrically 
         between the two PA molecules.  (b) : Lowest energy insulating structure 
         P1c  ($P1$ symmetry, stable). Here the two molecules underwent a disproportionation,  becoming 
         spontaneously inequivalent, one preferentially bound to two La ions and less planar, 
         the other more isolated and planar.} 
\label{FIG:STRUCTURE}
\end{figure}

After this full $P2_{1}$ search, a set of random $P1$ trial structures were created by 
intermixing the position of La atoms of the best high symmetry $P2_{1}$ structures, as 
follows. If two $P2_{1}$ structures $p$ and $q$ have the two La atoms at positions 
$(\bf{P}, \bf{P^{\prime}})$ and $(\bf{Q}, \bf{Q^{\prime}})$ respectively, a new trial structure, generally 
of $P1$ symmetry, is created by placing the two La atoms at $(\bf{P}, \bf{Q})$,  $(\bf{P}, \bf{Q^{\prime}})$,$(\bf{Q}, \bf{P})$, etc. 
Additional sets of low symmetry trial structures were created by high temperature 
{\sl ab-initio} molecular dynamics (MD) simulations\footnote{We are grateful to Prof.~R.~Martonak 
for his help with these simulations} (see Table~\ref{TAB:LANDSCAPE}). 
After that, newer attempts failed to yield radically new outcomes, and the variety of trial structures 
explored was deemed to be sufficient. 
DFT total energy and force calculation with subsequent optimization readily showed that several 
optimized $P1$ structures have a lower energy than the best $P2_{1}$ ones,  thus confirming the 
correctness of exploring reduced symmetries. Parameters of the few lowest energy relaxed  $P1$ 
structures are given in  Table~\ref{TAB:LANDSCAPE}. We note that they are close in energy and share 
a common feature: the two La atoms in the cell approach the same PA molecule,
rather than remaining equidistant between the two PA molecules as in the $P2_{1}$ structures. 
The two molecules display a slightly different shape, the isolated one  now
more planar than the other. An independent vibrational signature of this insulating 
phase could therefore consist of a characteristic splitting of the main intra-molecular 
vibrations reflecting the nonequivalence of the two PA molecules. 
Fig.~\ref{FIG:STRUCTURE}, shows schematically the lowest $P1$ 
structures, labeled P1c. 
The second lowest $P1$ structure obtained by 8512 MD-steps (called 8512s in 
Table~\ref{TAB:LANDSCAPE}) is nearly degenerate and isostructural with P1c.  Indeed, 
as clear in Fig.~\ref{FIG:ENERGYLAND}, all attempts, i.e. MD simulations with different 
initial structures, end up to insulating phases with nearly similar structures and energy. 
Among the higher symmetry $P2_{1}$ structures, 18 and 36 are the lowest in energy, again
nearly degenerate within our accuracy, and about 40 mRy per cell, that is more than 0.5 eV, higher
than the $P_1$ structures. The reason for this systematic energy lowering caused by disproportionation
is best understood by careful consideration of the electronic structures, which we will do next.

\begin{figure*}[htp!]
\centering
\begin{minipage}[c]{\linewidth}
      \begin{minipage}{0.42\linewidth}
      \includegraphics[width=1\linewidth]{FIG4A.eps}
      \end{minipage}
      \begin{minipage}{0.14\linewidth}
      \includegraphics[width=1\linewidth]{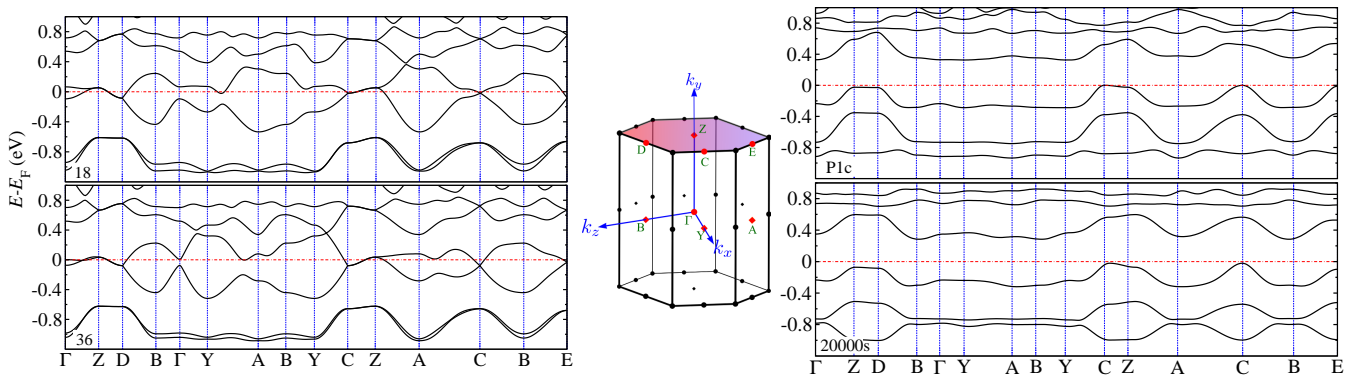}
      \end{minipage}
      \begin{minipage}{0.42\linewidth}
      \includegraphics[width=1\linewidth]{FIG4C.eps}
      \end{minipage}
\caption{(Color online) Electron band structures for the two best $P2_{1}$ structures 
         18 and 36, (metastable, metallic) and for the two best $P1$ structures P1c and 20000s (stable, insulating). 
         The Brillouin Zone is also shown, highlighting the 010 boundary plane where symmetry induced 
         band degeneracy occurs in the $P2_{1}$ structures.}
\label{FIG:BANDSDFT}
\end{minipage}
\end{figure*}

The electronic band structures of the optimized La-PA geometries are shown on Fig.\ref{FIG:BANDSDFT}. All the $P2_{1}$ structures 
are metallic, owing to a symmetry-induced degeneracy within the LUMO+1 derived band (nearest to 
the Fermi level) on the zone boundary plane $k_y= \pm \pi$, see Fig. \ref{FIG:BANDSDFT}.  
This is a symmetry-induced degeneracy where bands ``stick together" at zone boundary points,
a well known property of ``nonsymmorphic" space groups,\cite{heine60} groups that include a mixed 
rotation-fractional translation symmetry operation. The $P2_{1}$ symmetry includes a screw axis along $b$,
which makes the two PA molecules equivalent, while still distinct.  The band degeneracy of $P2_{1}$ La-PA 
near the Fermi level bears similarities with that reported for the metallic state of other doped PAHs like 
K$_3$-picene.\cite{subedi11, kosugi11, giovannetti11, Andres11,casula11}.
That invites a further pursuit of the present model $P2_{1}$ phase, even if only metastable
in La-PA, as a generic model of metallic doped PAH possibly prone to superconductivity. This is what 
we will do later in the following sections.  

The screw axis symmetry operation present in $P2_{1}$ is missing in the $P1$ structures, where the two 
LA molecules have become inequivalent. Because of that, the zone boundary 
degeneracy near $E_{\rm F}$ is lifted, and a band gap opens up which splits the two LUMO+1 derived bands
leading to an insulating state.  The $P1$ insulating state is thus closely connected with the 
valence disproportionation taking place at the structural level between the two PA molecules in the cell, 
one more closely approached, and partly covalently bonded, by the two La ions, the other more isolated. 
As seen in Fig.\ref{FIG:BANDSDFT} our calculated gap in La-PA is about 0.7 eV 
-- probably an underestimate, as is generally the case in (self-interaction uncorrected)  DFT. 
The striking parallelism of the valence and conduction bands further implies a strongly peaked 
joint density of states, suggesting large excitonic effects in optical absorbtion below the band 
gap.\cite{bassani75}  
We note that the present band insulating $P1$ state is a spin singlet, quite different from the 
antiferromagnetic band insulator state proposed for K$_3$-picene.\cite{giovannetti11}
Starting with the best $P2_{1}$ structure 18 of La-PA, we also searched for magnetic solutions of 
our DFT calculations. While we could not stabilize a ferromagnetic state, we did find a locally stable 
antiferromagnetic state, where the two molecules remain equivalent in both structure and electronic
charge, but exhibit oppositely polarized magnetizations of magnitude compatible with spin 1/2.  
However that solution was found for La-PA to lie only 0.7\,mRy lower that the parent $P2_1$ non-magnetic
metal, a minute energy gain by comparison with the ~ 40\,mRy gain of the nonmagnetic insulating 
$P1$ structure. \\ \indent   
Even if the $P2_1$ metallic structures are not of lowest energy, it is not unconceivable that they could 
still form and survive in a metastable state for kinetic reasons.
In the rest of the paper we will concentrate on these metallic $P2_{1}$ phases, in particular on structure 18,
as a possible, even if speculative, seat of superconductivity. Rather than addressing superconductivity at this stage, 
and notwithstanding the fact that some groups are currently 
pursuing electron-phonon coupling with direct {\sl ab-initio} methods,\cite{casula11,casula12} our alternative and 
present goal is to assemble in a single model what we believe could be the basic ingredients and elements 
for a simple scheme that would allow wider scope model superconductivity studies in the future. The elements we  
consider are of three kinds. The first is a minimal tight binding model electronic structure, including in 
this case only the two LUMO+1 derived bands. Similarly to earlier studies of K$_3$ picene,\cite{giovannetti11} 
we will do that for La-PA in the next section. 
As a second element we wish to identify at least one important lattice phonon that is strongly 
coupled to electrons near the Fermi level. We will argue in the subsequent section that symmetry 
suggests a ``dimerizing" phonon, whose eigenvector instantaneously turns $P2_1$ symmetry to $P1$,  
as a natural candidate. The third element, widely discussed e.g., in Ref \cite{giovannetti11, kosugi11} 
is the electron-electron intra-molecular repulsion $U$. Good estimates are already available,  
for phenanthrene and other PAHs triply negative ions.\cite{nomura12} In the next two chapters we 
thus focus on modeling the tight binding bands and the main electron-phonon coupling.

\section{Tight binding modeling of metallic bands}
\label{SEC:TBM}

In order to extract the hopping parameter between Wannier functions centered on the two equivalent
phenanthrene molecules ``1" and ``2" in the $P2_{1}$ cell, we implemented calculations using 
Wannier90~\cite{wannier90} and fitted the two LUMO+1 derived metallic bands of structure 18 
as highlighted by orange lines in Fig.~\ref{FIG:TOWBAND} 
restricting only to hopping matrix 
elements $\gtrsim 0.01$\,eV. 

\begin{figure*}[htp!]
\centering
\includegraphics[width=1.\linewidth]{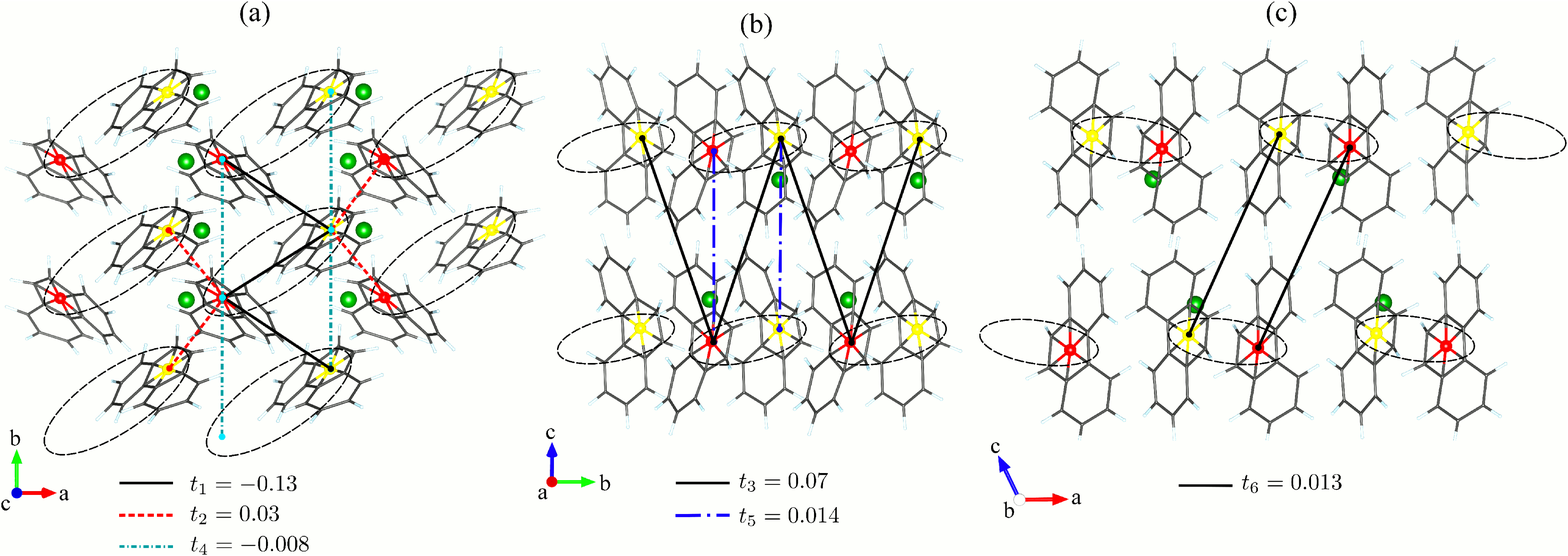}
\caption{(Color online) Wannier function hopping parameters in $a-b$, $b-c$, and $a-c$ planes. 
         Red dot: PA molecule 1; yellow dot: PA molecule 2; green dot: La atom. 
The unit cell containing two PA molecules is encircled by dashed lines.
         The strongest hopping elements are drawn by heavy lines.}
\label{FIG:abplane}
\end{figure*}

Fig.~\ref{FIG:abplane} (a) shows the $ab$-plane with the strongest hopping elements. 
Molecules ``1" and ``2" in the origin unit cell have coordinates  $(0,0,0)$ and $\frac{1}{2}\,(1,1,0)$, respectively.   
We note that the molecular staggering brings about hopping amplitudes that are not 
invariant under $a\to -a$. The hopping matrix $\hat{t}_\bk$ with elements $t^{ij}_\bk$, $i,j=1,2$ reads 
\be
\hat{t}_\bk = 
\begin{pmatrix}
t^{11}_\bk & t^{12}_\bk \\
~ & ~\\
t^{21}_\bk & t^{22}_\bk
\end{pmatrix}, \label{t-matrix}
\ee
where 
\ba
t^{12}_\bk &=& \bigg(1+\text{e}^{-i\bk\cdot \mathbf{b}}\bigg)\;\Bigg(
t_1 + t_2\,\text{e}^{-i\bk\cdot \mathbf{a}}
 + t_3\,\text{e}^{i\bk\cdot \mathbf{c}}\Bigg) \equiv -\text{e}^{-i\theta_\bk}\; \tau_\bk,
\ea
$t^{21}_\bk = t^{12\,*}_\bk$, and   
\be
\tau_\bk = \mid t^{12}_\bk\mid = 
2\left|\cos\left(\frac{\bk\cdot \mathbf{b}}{2}\right)\right| \,
\bigg|t_1 + t_2\,\text{e}^{-i\bk\cdot \mathbf{a}}
 + t_3\,\text{e}^{i\bk\cdot \mathbf{c}}\bigg|, \label{tau-k}
\ee
while
\be
t^{11}_\bk=t^{22}_\bk \equiv t_\bk = 2t_4\cos \bk\cdot \mathbf{b} + 2t_5 \cos \bk\cdot \mathbf{c}
+ 2t_6\cos \bk\cdot\big(\mathbf{a}+\mathbf{c}\big) \nonumber.
\ee
The two bands have therefore a dispersion 
\be
\epsilon_{\pm \bk} = t_\bk \pm \tau_\bk. \label{e-k}
\ee
These model bands are degenerate on the whole plane $\bk\cdot\mathbf{b}= \pi$, where  
in addition they disperse very little, see Fig. 4. 
The optimized hopping parameters are;  $t_1$ = -0.13, $t_2$ = 0.03, $t_3$ = 0.07, 
$t_4$ = -0.008, $t_5$ =  0.014, $t_6$ = 0.013 (see Fig.~\ref{FIG:abplane}). 

We note that the degeneracy on the plane $\bk\cdot\mathbf{b}= \pi$ can be easily lifted   
once the screw axis along $b$,  a symmetry present in $P2_{1}$ structures, is removed. 
The removal can occur in different ways. The two molecules can cease to be equivalent as in
the disproportionated $P1$ phase described above. Then  
$t^{11}_\bk\not = t^{22}_\bk $ in Eq. \eqn{t-matrix}. In particular, if $t^{11}_\bk-t^{22}_\bk = \Delta$ is 
the LUMO+1 orbital splitting of the two molecules in the unit cell, then $\tau_\bk$ in Eq. \eqn{tau-k} changes into 
\be
\tau_\bk \rightarrow \sqrt{ \Delta^2 + \left|t^{12}_\bk\right|^2},
\ee
so that the degeneracy is lost at $\bk\cdot\mathbf{b}=\pi$, where $\left|t^{12}_\bk\right|^2=0$. 
Another possibility to remove the screw axis induced degeneracy is dimerization of PA molecules, 
even without inequivalence. Let us assume, for instance, 
that the hopping $t_1$ between two molecules in the same unit cell increases $t_1\to t_1(1+\delta)$, while 
that between molecules on different unit cells diminishes $t_1\to t_1(1-\delta)$. It follows that 
\be
t^{12}_\bk \rightarrow t^{12}_\bk + t_1\delta \bigg(1-\text{e}^{-i\bk\cdot \mathbf{b}}\bigg),
 \ee
hence 
\be
\tau_\bk \rightarrow \sqrt{4t_1^2\delta^2\sin^2\left(\frac{\bk\cdot \mathbf{b}}{2}\right)
+ \left|t^{12}_\bk\right|^2},
\ee
does not vanish anymore on the plane $\bk\cdot\mathbf{b}= \pi$. 

It is clear that if the energy splitting $\Delta$ or the dimerization $\delta$ exceed a threshold, then the 
two LUMO+1 derived bands cease to overlap and the band structure turns insulating, 
which is impossible if the two bands remains strictly degenerate on the plane $\bk\cdot\mathbf{b}= \pi$.
We finally note, as mentioned above, that an insulating state could also be realized by a sufficiently strong antiferromagnetic ordering, inducing a 
spin dependent splitting $\Delta_\up = - \Delta_\down$ between the LUMO+1 orbitals of the two molecules. 

Thus the metallic state is highly fragile against perturbations or fluctuations that either tend to make the two 
molecules inequivalent, with or without magnetism, or else induce a 
dimerization pattern. If we aim at identifying processes that could mediate 
superconductivity, it is a natural choice to focus on either structural (phonon) fluctuations, or on
antiferromagnetic fluctuations of the above types, since they would  most readily destabilize the metallic 
state. Although we cannot exclude that magnetic fluctuations could play the major 
role, we shall follow here a more conventional BCS approach and analyze qualitatively 
the electron-phonon coupling to modes that make the metallic state most unstable.

\section{ Dimerizing Phonon}
\label{SEC:DIMER}

We want to address and model here the simplest and most basic mechanism  by which the 
metallic electronic states near the Fermi level are coupled to vibrations of the La-PA lattice.
As an alternative to full fledge, system specific DFT electron phonon calculations which already exist ,\cite{casula11} 
we wish to pursue here a simplified model of more generic use. We are guided by the physical consideration that the most 
important  vibrations are all those that instantaneously break the symmetry induced degeneracy 
and open up a gap near the Fermi level. In that sense, the
situation bears some analogy with that of MgB$_2$, where a frozen-in E$_{2g}$ vibration linearly
splits the degeneracy of the $\sigma$ band edge at the $\Gamma$ point and near the Fermi level.\cite{shukla03}

\begin{figure} [htb!]
\centering
\includegraphics[width=1.\linewidth]{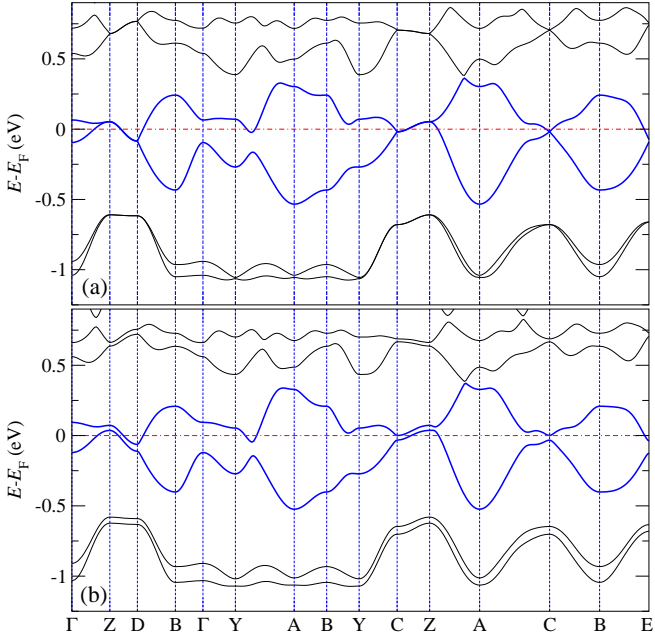}
\caption{(Color online) Band structure of metallic La-PA (structure 18)  (a) before and (b) after a frozen-in
dimerization distortion modulating the distance of molecules 1 and 2 by $\pm$ 0.08\,\AA 
in the $ab$ plane,
along the direction of hopping $t_{1}$ (see Fig.~\ref{FIG:abplane}).
The frozen phonon very effectively lifts the 
symmetry induced degeneracy at the  Brillouin zone boundary opening up a gap at the Fermi level, 
see blue lines).}
\label{FIG:Dimerization}
\end{figure}

In our $P2_{1}$ model structure, the degeneracy on the plane $\bk\cdot\mathbf{b}=\pi$ can be removed by 
 $\bq= 0$ optical phonon, of either 
 intramolecular or intermolecular character, inducing, respectively, an instantaneous energy splitting $\Delta$ 
 or dimerization $\delta$, as previously discussed. Although both phonons can be expected to be relevant 
 in practice, as was found in electron phonon calculations of K$_3$ picene\cite{casula12}
 we focus our attention on the intermolecular mode.
 The strongest intermolecular hopping is $t_{1}$ in the 
 $ab$-plane. (see Fig.~\ref{FIG:abplane}).
A lattice vibration bringing closer the two molecules within each unit cell along the direction of hopping $t_1$ is therefore a reasonable first guess for a strongly coupled mode. 

We thence consider a lattice periodic static distortion 
-- a frozen phonon --  corresponding to the above mentioned dimerization, where molecules 1 and 2 
within each unit cell move toward each other in the $ab$-plane. 
We showed previously that this
frozen phonon can destroy the zone boundary degeneracy, and open very effectively a band gap. 
To verify the magnitude of that effect we carried out DFT calculations where the initial metallic 
structure 18 was progressively ``dimerized" in steps of 0.02\,\AA. Fig.~\ref{FIG:Dimerization} 
compares the band structure of original La-PA 18 with that where the 1-2-1 distances in the $ab$ 
plane now alternate by  $\pm$ 0.08\,\AA.  As is seen, the gap indeed opens with great efficiency, 
whereas the rest of bands remain practically intact. 

It is worth highlighting that the dimerizing distortion,  
although able to split the degeneracy on the $\mathbf{k}\cdot\mathbf{b}=\pi$ plane, nonetheless 
increases the total energy. In other words, the original undistorted $P2_{1}$ structure is stable towards
a dimerization that leaves the two molecules within the unit cell equivalent, as well as towards a weak distortion that induces charge disproportionation between the two molecules, thus making them not equivalent.  In fact, the  $P2_{1}$ structure is found to be a stable local minimum, disconnected from the absolute minimum of $P1$ symmetry previously discussed. 
\subsection{Effective two-band model}
\label{SUBSEC:MODEL}

We now include the dimerization mode in a model Hamiltonian for the LUMO+1 derived bands. 
The non-interacting tight-binding Hamiltonian is that of section \ref{SEC:TBM}, i.\,e. 
\begin{figure} [htb!]
\centering
\includegraphics[width=1.\linewidth]{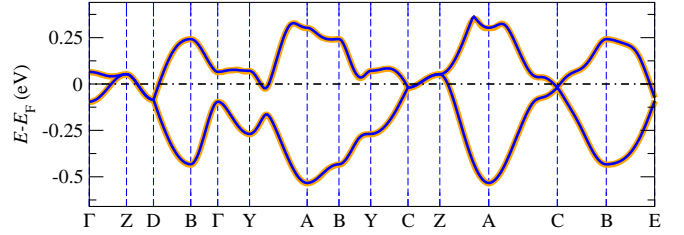}
\caption{(Color online)  The two  LUMO+1 derived bands used for modeling. 
         The bold orange lines are Wannier parametrized bands, indistinguishable from the DFT results. 
         Extracted  hopping parameters are listed in Fig~\ref{FIG:abplane}.}
\label{FIG:TOWBAND}
\end{figure}

\be
\mathcal{H}_0 = \sum_{\bk\sigma}\, 
\Big(c^\dagger_{1\bk\sigma},c^\dagger_{2\bk\sigma}\Big)
\begin{pmatrix}
t_\bk & t^{12}_\bk\\
t^{12\,*}_\bk & t_\bk
\end{pmatrix}
\begin{pmatrix}
c^\dagga_{1\bk\sigma}\\
c^\dagga_{2\bk\sigma}
\end{pmatrix},
\ee
where $c^\dagger_{1(2)\bk\sigma}$ creates an electron in LUMO+1 orbital 
(see Fig.~\ref{FIG:TOWBAND}) of molecule $1(2)$ with crystal momentum $\bk$ 
and spin $\sigma$. We assume that each molecule can displace from its equilibrium 
position and such displacement affects primarily the hopping between molecule 1 and 2. 
Specifically, we shall consider the electron-phonon coupling of the classic 
``Fr{\"o}hlich" type ~\cite{Frol1_50,Frol2_50}
\bea
\mathcal{H}_\text{el-ph} &=& 
\sum_{\bR,\bR_i}\sum_\sigma \,
g_{|\bR-\boldsymbol{\delta}|}\,t^{12}_\bR\,\big(\mathbf{u}_{1\bR_i+\bR}-\mathbf{u}_{2\bR_i}\big)\cdot\big(\bR-\boldsymbol{\delta}\big)
\nonumber\\
&&~~~~~~~~~~\Big(c^\dagger_{1\bR+\bR_i\sigma}c^\dagga_{2\bR_i\sigma}+H.c.\Big),
\label{H-el-ph}
\eea 
where $\mathbf{u}_{1(2)\bR}$ is the displacement of the molecule 1(2) in the unit cell with coordinate 
$\bR$ and we have assumed that the variation of the hopping at linear order in the displacement,
\[
\delta t^{12}_\bR \simeq g_{|\bR-\boldsymbol{\delta}|}\,t^{12}_\bR\,\big(\mathbf{u}_{1\bR_i+\bR}-\mathbf{u}_{2\bR_i}\big)\cdot\big(\bR-\boldsymbol{\delta}\big),
\]
is proportional to $t^{12}_\bR$ at the equilibrium positions, with $\boldsymbol{\delta}$ 
the vector connecting the two molecules in the unit cell. We shall concentrate on 
displacement modes polarized along the $b$ axis, i.e. $\mathbf{u}_{1(2)\bR} = \big(u^a_{1(2)\bR}, u^b_{1(2)\bR},u^c_{1(2)\bR}\big) 
= \big(0, u_{1(2)\bR},0\big) $, 
which can induce a dimerization pattern able to turn the metal into an insulator. It follows that the 
Fourier transform of the electron-phonon coupling
\be
\gamma_\bk \equiv \sum_\bR\, g_{|\bR-\boldsymbol{\delta}|}\,t^{12}_\bR\, 
\text{e}^{-i\bk\cdot\bR}\; 
\big(R^b-\delta^b\big)  \propto \Big(1-\text{e}^{-i\bk\cdot\mathbf{b}}\Big),\label{gamma-k}
\ee
hence finite on the degeneracy plane $\bk\cdot\mathbf{b}= \pi$. The Fourier transform of 
the electron-phonon Hamiltonian \eqn{H-el-ph} thus reads
\bea
&& \mathcal{H}_\text{el-ph} = \sum_{\bk,\bq,\sigma}\, 
\Big(u_{1-\bq}\,\gamma_{\bk+\bq} - u_{2-\bq}\,\gamma_\bk\Big)\,
c^\dagger_{1\bk\sigma} c^\dagga_{2\bk+\bq\sigma}\nonumber \\
&&~~~~~~~~ + \Big(u_{1-\bq}\,\gamma_{-\bk} - u_{2-\bq}\,\gamma_{-\bk-\bq}\Big)\,
c^\dagger_{2\bk\sigma} c^\dagga_{1\bk+\bq\sigma}. 
\eea
An out-of-phase displacement of the two molecules is characterized by $u_{1\bq} = x_\bq$ 
and $u_{2\bq} = - \text{e}^{-i\bq\cdot\mathbf{b}/2}\;x_\bq$, with $x_\bq$ the optical eigenmode, so that 
\bea
&& \mathcal{H}_\text{el-ph} = \sum_{\bk,\bq,\sigma}\, x_{-\bq}
\Big(\gamma_{\bk+\bq} + \text{e}^{i\bq\cdot\mathbf{b}/2}\;\gamma_\bk\Big)\,
c^\dagger_{1\bk\sigma} c^\dagga_{2\bk+\bq\sigma}\nonumber \\
&&~~~~~~~~ + x_{-\bq}\Big(\gamma_{-\bk} + \text{e}^{i\bq\cdot\mathbf{b}/2}\;\gamma_{-\bk-\bq}\Big)\,
c^\dagger_{2\bk\sigma} c^\dagga_{1\bk+\bq\sigma}.\label{H-el-ph-final} 
\eea

In addition, we must include the phonon Hamiltonian 
\be
\mathcal{H}_\text{ph} = \sum_\bq \,\frac{\omega_\bq}{2}\,\Big(p_\bq p_{-\bq} + x_\bq x_{-\bq}\Big),
\label{H-ph}
\ee
with $\big[x_{\bq},p_{-\bq'}\big] = i\delta_{\bq\bq'}$.

The final ingredient we need to include is the electron-electron interaction. The simplest term that accounts 
for the suppression of charge fluctuations in such a narrow-band molecular conductor is a Hubbard repulsion
\be
\mathcal{H}_\text{U} = U\, \sum_\bR\, n_{1\bR\up}n_{1\bR\down} + n_{2\bR\up}n_{2\bR\down}.
\label{H-U}
\ee

Numerical values for $U$ have been given for a number of PAHs including phenanthrene by Nomura et al.~\cite{nomura12}
They were generally found to be large, in fact quite similar to the electron bandwidths. 
In reality, since those values refer to a 4-band model that includes also LUMO besides LUMO+1, 
the corresponding estimates for a 2-band model aiming at describing just LUMO+1 must be smaller, since  they already account for the LUMO screening of the effective $U$ among LUMO+1 electrons. 

In summary, our proposed Hamiltonian is a ``Hubbard-Fr{\"o}hlich" one  
\be
\mathcal{H} = \mathcal{H}_0 + \mathcal{H}_\text{ph}+\mathcal{H}_\text{el-ph} 
+ \mathcal{H}_\text{U},\label{Ham}
\ee
It does in our view include the main ingredients present in  
a generic $P2_1$ metallic phase -- La-PA, or in fact any other three-electron-doped PAH crystal with
that kind of high symmetry.  

\section{Discussion and Conclusions}
\label{SEC:CONC}

We have conducted a theoretical study of La-phenanthrene, here adopetd as a generic model of 
electron-doped PAHs, where a number of superconducting phases have been experimentally proposed.
Exploring by ab initio calculations a large variety of bimolecular structures, we found that the lowest energy state
of La-PA is not metallic as expected, but is a band insulator resulting from a symmetry-lowering disproportionation of the
two PA molecules. In that phase of La-PA an optical band gap of 0.7 eV\ or larger should be observable in near infrared,
along with vibrational signatures of the disproportionation between the two molecules.

At higher total energy we also found and characterized high symmetry metallic phases, whose electronic structures
resemble those proposed for alkali doped picene. Although in La-PA this metallic phase could only exist as  a 
metastable state, we studied it in detail with the scope of extracting a generic two-band model containing 
interesting elements that could determine and control the superconductivity of more general electron-doped PAHs, where
many elements are similar. The two bands are derived from LUMO+1 molecular orbitals and were extracted from a tight binding 
fit of the La-PA ab initio bands near the Fermi level. An important phonon with a standard Fr{\"o}hlich coupling to the 
electrons in these bands  was identified as a dimerizing intermolecular vibration, which acts by breaking a 
symmetry-induced band degeneracy near Fermi. The strong electron-electron repulsion $U$ typical of 
PAH ions constitutes  the third important element of that model. We conclude proposing this overall 
two band  ``Hubbard-Fr{\"o}hlich" model for further studies of superconductivity in these systems. While the 
superconducting solution of this model and its properties will be the subject of subsequent work,
we underline here that it should in principle be applicable, given its symmetry motivated, highly schematic form, 
to a broad family of electron-doped PAHs, and not just to La-PA,  whose metallic phase, predicted to be
metastable at best, is still waiting  to be properly identified and characterized.

\begin{acknowledgments}
This work was support by the European Union FP7-NMP-2011-EU-Japan 
project LEMSUPER. We acknowledge discussions with, and relevant information from, Y. Kubozono, H. Aoki, T. Kariyado, 
G. Giovannetti, M. Capone, and P. Carretta. We acknowledge the CINECA award 2013 for the availability of 
high performance computing resources and support.
\end{acknowledgments}


\begin{thebibliography}{42}%
\makeatletter
\providecommand \@ifxundefined [1]{%
 \@ifx{#1\undefined}
}%
\providecommand \@ifnum [1]{%
 \ifnum #1\expandafter \@firstoftwo
 \else \expandafter \@secondoftwo
 \fi
}%
\providecommand \@ifx [1]{%
 \ifx #1\expandafter \@firstoftwo
 \else \expandafter \@secondoftwo
 \fi
}%
\providecommand \natexlab [1]{#1}%
\providecommand \enquote  [1]{``#1''}%
\providecommand \bibnamefont  [1]{#1}%
\providecommand \bibfnamefont [1]{#1}%
\providecommand \citenamefont [1]{#1}%
\providecommand \href@noop [0]{\@secondoftwo}%
\providecommand \href [0]{\begingroup \@sanitize@url \@href}%
\providecommand \@href[1]{\@@startlink{#1}\@@href}%
\providecommand \@@href[1]{\endgroup#1\@@endlink}%
\providecommand \@sanitize@url [0]{\catcode `\\12\catcode `\$12\catcode
  `\&12\catcode `\#12\catcode `\^12\catcode `\_12\catcode `\%12\relax}%
\providecommand \@@startlink[1]{}%
\providecommand \@@endlink[0]{}%
\providecommand \url  [0]{\begingroup\@sanitize@url \@url }%
\providecommand \@url [1]{\endgroup\@href {#1}{\urlprefix }}%
\providecommand \urlprefix  [0]{URL }%
\providecommand \Eprint [0]{\href }%
\providecommand \doibase [0]{http://dx.doi.org/}%
\providecommand \selectlanguage [0]{\@gobble}%
\providecommand \bibinfo  [0]{\@secondoftwo}%
\providecommand \bibfield  [0]{\@secondoftwo}%
\providecommand \translation [1]{[#1]}%
\providecommand \BibitemOpen [0]{}%
\providecommand \bibitemStop [0]{}%
\providecommand \bibitemNoStop [0]{.\EOS\space}%
\providecommand \EOS [0]{\spacefactor3000\relax}%
\providecommand \BibitemShut  [1]{\csname bibitem#1\endcsname}%
\let\auto@bib@innerbib\@empty
\bibitem [{\citenamefont {Mitsuhashi}\ \emph {et~al.}(2010)\citenamefont
  {Mitsuhashi}, \citenamefont {Suzuki}, \citenamefont {Yamanari}, \citenamefont
  {Mitamura}, \citenamefont {Kambe}, \citenamefont {Ikeda}, \citenamefont
  {Okamoto}, \citenamefont {Fujiwara}, \citenamefont {Yamaji}, \citenamefont
  {Kawasaki}, \citenamefont {Maniwa},\ and\ \citenamefont
  {and}}]{mitsuhashi10}%
  \BibitemOpen
  \bibfield  {author} {\bibinfo {author} {\bibfnamefont {R.}~\bibnamefont
  {Mitsuhashi}}, \bibinfo {author} {\bibfnamefont {Y.}~\bibnamefont {Suzuki}},
  \bibinfo {author} {\bibfnamefont {Y.}~\bibnamefont {Yamanari}}, \bibinfo
  {author} {\bibfnamefont {H.}~\bibnamefont {Mitamura}}, \bibinfo {author}
  {\bibfnamefont {T.}~\bibnamefont {Kambe}}, \bibinfo {author} {\bibfnamefont
  {N.}~\bibnamefont {Ikeda}}, \bibinfo {author} {\bibfnamefont
  {H.}~\bibnamefont {Okamoto}}, \bibinfo {author} {\bibfnamefont
  {A.}~\bibnamefont {Fujiwara}}, \bibinfo {author} {\bibfnamefont
  {M.}~\bibnamefont {Yamaji}}, \bibinfo {author} {\bibfnamefont
  {N.}~\bibnamefont {Kawasaki}}, \bibinfo {author} {\bibfnamefont
  {Y.}~\bibnamefont {Maniwa}}, \ and\ \bibinfo {author} {\bibfnamefont
  {Y.}~\bibnamefont {and}},\ }\href {\doibase 10.1038/nature08859} {\bibfield
  {journal} {\bibinfo  {journal} {Nature}\ }\textbf {\bibinfo {volume} {464}},\
  \bibinfo {pages} {76} (\bibinfo {year} {2010})}\BibitemShut {NoStop}%
\bibitem [{\citenamefont {Kubozono}\ \emph {et~al.}(2011)\citenamefont
  {Kubozono}, \citenamefont {Mitamura}, \citenamefont {Lee}, \citenamefont
  {He}, \citenamefont {Yamanari}, \citenamefont {Takahashi}, \citenamefont
  {Suzuki}, \citenamefont {Kaji}, \citenamefont {Eguchi}, \citenamefont
  {Akaike}, \citenamefont {Kambe}, \citenamefont {Okamoto}, \citenamefont
  {Fujiwara}, \citenamefont {Kato}, \citenamefont {Kosugi},\ and\ \citenamefont
  {Aoki}}]{kubozono11}%
  \BibitemOpen
  \bibfield  {author} {\bibinfo {author} {\bibfnamefont {Y.}~\bibnamefont
  {Kubozono}}, \bibinfo {author} {\bibfnamefont {H.}~\bibnamefont {Mitamura}},
  \bibinfo {author} {\bibfnamefont {X.}~\bibnamefont {Lee}}, \bibinfo {author}
  {\bibfnamefont {X.}~\bibnamefont {He}}, \bibinfo {author} {\bibfnamefont
  {Y.}~\bibnamefont {Yamanari}}, \bibinfo {author} {\bibfnamefont
  {Y.}~\bibnamefont {Takahashi}}, \bibinfo {author} {\bibfnamefont
  {Y.}~\bibnamefont {Suzuki}}, \bibinfo {author} {\bibfnamefont
  {Y.}~\bibnamefont {Kaji}}, \bibinfo {author} {\bibfnamefont {R.}~\bibnamefont
  {Eguchi}}, \bibinfo {author} {\bibfnamefont {K.}~\bibnamefont {Akaike}},
  \bibinfo {author} {\bibfnamefont {T.}~\bibnamefont {Kambe}}, \bibinfo
  {author} {\bibfnamefont {H.}~\bibnamefont {Okamoto}}, \bibinfo {author}
  {\bibfnamefont {A.}~\bibnamefont {Fujiwara}}, \bibinfo {author}
  {\bibfnamefont {T.}~\bibnamefont {Kato}}, \bibinfo {author} {\bibfnamefont
  {T.}~\bibnamefont {Kosugi}}, \ and\ \bibinfo {author} {\bibfnamefont
  {H.}~\bibnamefont {Aoki}},\ }\href {\doibase 10.1039/C1CP20961B} {\bibfield
  {journal} {\bibinfo  {journal} {Phys.~Chem.~Chem.~Phys.}\ }\textbf {\bibinfo
  {volume} {13}},\ \bibinfo {pages} {16476} (\bibinfo {year}
  {2011})}\BibitemShut {NoStop}%
\bibitem [{\citenamefont {Kato}\ \emph {et~al.}(2011)\citenamefont {Kato},
  \citenamefont {Kambe},\ and\ \citenamefont {Kubozono}}]{kato11}%
  \BibitemOpen
  \bibfield  {author} {\bibinfo {author} {\bibfnamefont {T.}~\bibnamefont
  {Kato}}, \bibinfo {author} {\bibfnamefont {T.}~\bibnamefont {Kambe}}, \ and\
  \bibinfo {author} {\bibfnamefont {Y.}~\bibnamefont {Kubozono}},\ }\href
  {\doibase 10.1103/PhysRevLett.107.077001} {\bibfield  {journal} {\bibinfo
  {journal} {Phys.~Rev.~Lett.}\ }\textbf {\bibinfo {volume} {107}},\ \bibinfo
  {pages} {077001} (\bibinfo {year} {2011})}\BibitemShut {NoStop}%
\bibitem [{\citenamefont {Xue}\ \emph {et~al.}(2012)\citenamefont {Xue},
  \citenamefont {Cao}, \citenamefont {Wang}, \citenamefont {Wu}, \citenamefont
  {Yang}, \citenamefont {Dong}, \citenamefont {He}, \citenamefont {Li},\ and\
  \citenamefont {Chen}}]{xue12}%
  \BibitemOpen
  \bibfield  {author} {\bibinfo {author} {\bibfnamefont {M.}~\bibnamefont
  {Xue}}, \bibinfo {author} {\bibfnamefont {T.}~\bibnamefont {Cao}}, \bibinfo
  {author} {\bibfnamefont {D.}~\bibnamefont {Wang}}, \bibinfo {author}
  {\bibfnamefont {Y.}~\bibnamefont {Wu}}, \bibinfo {author} {\bibfnamefont
  {H.}~\bibnamefont {Yang}}, \bibinfo {author} {\bibfnamefont {X.}~\bibnamefont
  {Dong}}, \bibinfo {author} {\bibfnamefont {J.}~\bibnamefont {He}}, \bibinfo
  {author} {\bibfnamefont {F.}~\bibnamefont {Li}}, \ and\ \bibinfo {author}
  {\bibfnamefont {G.~F.}\ \bibnamefont {Chen}},\ }\href {\doibase
  10.1038/srep00389} {\bibfield  {journal} {\bibinfo  {journal} {Sci.~Rep.}\
  }\textbf {\bibinfo {volume} {2}},\ \bibinfo {pages} {389} (\bibinfo {year}
  {2012})}\BibitemShut {NoStop}%
\bibitem [{\citenamefont {Wang}\ \emph
  {et~al.}(2011{\natexlab{a}})\citenamefont {Wang}, \citenamefont {Liu},
  \citenamefont {Gui}, \citenamefont {Xie}, \citenamefont {Yan}, \citenamefont
  {Ying}, \citenamefont {Luo},\ and\ \citenamefont {Chen}}]{wang11}%
  \BibitemOpen
  \bibfield  {author} {\bibinfo {author} {\bibfnamefont {X.}~\bibnamefont
  {Wang}}, \bibinfo {author} {\bibfnamefont {R.}~\bibnamefont {Liu}}, \bibinfo
  {author} {\bibfnamefont {Z.}~\bibnamefont {Gui}}, \bibinfo {author}
  {\bibfnamefont {Y.}~\bibnamefont {Xie}}, \bibinfo {author} {\bibfnamefont
  {Y.}~\bibnamefont {Yan}}, \bibinfo {author} {\bibfnamefont {J.}~\bibnamefont
  {Ying}}, \bibinfo {author} {\bibfnamefont {X.}~\bibnamefont {Luo}}, \ and\
  \bibinfo {author} {\bibfnamefont {X.}~\bibnamefont {Chen}},\ }\href {\doibase
  10.1038/ncomms1513} {\bibfield  {journal} {\bibinfo  {journal} {Nature
  Commun.}\ }\textbf {\bibinfo {volume} {2}},\ \bibinfo {pages} {507} (\bibinfo
  {year} {2011}{\natexlab{a}})}\BibitemShut {NoStop}%
\bibitem [{\citenamefont {Wang}\ \emph
  {et~al.}(2011{\natexlab{b}})\citenamefont {Wang}, \citenamefont {Yan},
  \citenamefont {Gui}, \citenamefont {Liu}, \citenamefont {Ying}, \citenamefont
  {Luo},\ and\ \citenamefont {Chen}}]{wang11a}%
  \BibitemOpen
  \bibfield  {author} {\bibinfo {author} {\bibfnamefont {X.~F.}\ \bibnamefont
  {Wang}}, \bibinfo {author} {\bibfnamefont {Y.~J.}\ \bibnamefont {Yan}},
  \bibinfo {author} {\bibfnamefont {Z.}~\bibnamefont {Gui}}, \bibinfo {author}
  {\bibfnamefont {R.~H.}\ \bibnamefont {Liu}}, \bibinfo {author} {\bibfnamefont
  {J.~J.}\ \bibnamefont {Ying}}, \bibinfo {author} {\bibfnamefont {X.~G.}\
  \bibnamefont {Luo}}, \ and\ \bibinfo {author} {\bibfnamefont {X.~H.}\
  \bibnamefont {Chen}},\ }\href {\doibase 10.1103/PhysRevB.84.214523}
  {\bibfield  {journal} {\bibinfo  {journal} {Phys. Rev. B}\ }\textbf {\bibinfo
  {volume} {84}},\ \bibinfo {pages} {214523} (\bibinfo {year}
  {2011}{\natexlab{b}})}\BibitemShut {NoStop}%
\bibitem [{\citenamefont {Wang}\ \emph {et~al.}(2012)\citenamefont {Wang},
  \citenamefont {Luo}, \citenamefont {Ying}, \citenamefont {Xiang},
  \citenamefont {Zhang}, \citenamefont {Zhang}, \citenamefont {Zhang},
  \citenamefont {Yan}, \citenamefont {Wang}, \citenamefont {Cheng},
  \citenamefont {Ye},\ and\ \citenamefont {Chen}}]{wang12}%
  \BibitemOpen
  \bibfield  {author} {\bibinfo {author} {\bibfnamefont {X.~F.}\ \bibnamefont
  {Wang}}, \bibinfo {author} {\bibfnamefont {X.~G.}\ \bibnamefont {Luo}},
  \bibinfo {author} {\bibfnamefont {J.~J.}\ \bibnamefont {Ying}}, \bibinfo
  {author} {\bibfnamefont {Z.~J.}\ \bibnamefont {Xiang}}, \bibinfo {author}
  {\bibfnamefont {S.~L.}\ \bibnamefont {Zhang}}, \bibinfo {author}
  {\bibfnamefont {R.~R.}\ \bibnamefont {Zhang}}, \bibinfo {author}
  {\bibfnamefont {Y.~H.}\ \bibnamefont {Zhang}}, \bibinfo {author}
  {\bibfnamefont {Y.~J.}\ \bibnamefont {Yan}}, \bibinfo {author} {\bibfnamefont
  {A.~F.}\ \bibnamefont {Wang}}, \bibinfo {author} {\bibfnamefont
  {P.}~\bibnamefont {Cheng}}, \bibinfo {author} {\bibfnamefont {G.~J.}\
  \bibnamefont {Ye}}, \ and\ \bibinfo {author} {\bibfnamefont {X.~H.}\
  \bibnamefont {Chen}},\ }\href {\doibase 10.1088/0953-8984/24/34/345701}
  {\bibfield  {journal} {\bibinfo  {journal} {J.~Phys.~Condens.~Matter}\
  }\textbf {\bibinfo {volume} {24}},\ \bibinfo {pages} {345701} (\bibinfo
  {year} {2012})}\BibitemShut {NoStop}%
\bibitem [{\citenamefont {Mahns}\ \emph {et~al.}(2012)\citenamefont {Mahns},
  \citenamefont {Roth},\ and\ \citenamefont {Knupfer}}]{mahns12}%
  \BibitemOpen
  \bibfield  {author} {\bibinfo {author} {\bibfnamefont {B.}~\bibnamefont
  {Mahns}}, \bibinfo {author} {\bibfnamefont {F.}~\bibnamefont {Roth}}, \ and\
  \bibinfo {author} {\bibfnamefont {M.}~\bibnamefont {Knupfer}},\ }\href
  {\doibase 10.1063/1.3699188} {\bibfield  {journal} {\bibinfo  {journal}
  {J.~Chem.~Phys.}\ }\textbf {\bibinfo {volume} {136}},\ \bibinfo {pages}
  {134503} (\bibinfo {year} {2012})}\BibitemShut {NoStop}%
\bibitem [{\citenamefont {Caputo}\ \emph {et~al.}(2012)\citenamefont {Caputo},
  \citenamefont {Santo}, \citenamefont {Parisse}, \citenamefont {Petaccia},
  \citenamefont {Floreano}, \citenamefont {Verdini}, \citenamefont {Panighel},
  \citenamefont {Struzzi}, \citenamefont {Taleatu}, \citenamefont {Lal}, ,\
  and\ \citenamefont {Goldoni}}]{caputo12}%
  \BibitemOpen
  \bibfield  {author} {\bibinfo {author} {\bibfnamefont {M.}~\bibnamefont
  {Caputo}}, \bibinfo {author} {\bibfnamefont {G.~D.}\ \bibnamefont {Santo}},
  \bibinfo {author} {\bibfnamefont {P.}~\bibnamefont {Parisse}}, \bibinfo
  {author} {\bibfnamefont {L.}~\bibnamefont {Petaccia}}, \bibinfo {author}
  {\bibfnamefont {L.}~\bibnamefont {Floreano}}, \bibinfo {author}
  {\bibfnamefont {A.}~\bibnamefont {Verdini}}, \bibinfo {author} {\bibfnamefont
  {M.}~\bibnamefont {Panighel}}, \bibinfo {author} {\bibfnamefont
  {C.}~\bibnamefont {Struzzi}}, \bibinfo {author} {\bibfnamefont
  {B.}~\bibnamefont {Taleatu}}, \bibinfo {author} {\bibfnamefont
  {C.}~\bibnamefont {Lal}}, , \ and\ \bibinfo {author} {\bibfnamefont
  {A.}~\bibnamefont {Goldoni}},\ }\href {\doibase 10.1021/jp306640z} {\bibfield
   {journal} {\bibinfo  {journal} {J.~Phys.~Chem.~C}\ }\textbf {\bibinfo
  {volume} {116}},\ \bibinfo {pages} {19902} (\bibinfo {year}
  {2012})}\BibitemShut {NoStop}%
\bibitem [{\citenamefont {Subedi}\ and\ \citenamefont
  {Boeri}(2011)}]{subedi11}%
  \BibitemOpen
  \bibfield  {author} {\bibinfo {author} {\bibfnamefont {A.}~\bibnamefont
  {Subedi}}\ and\ \bibinfo {author} {\bibfnamefont {L.}~\bibnamefont {Boeri}},\
  }\href {\doibase 10.1103/PhysRevB.84.020508} {\bibfield  {journal} {\bibinfo
  {journal} {Phys.~Rev.~B}\ }\textbf {\bibinfo {volume} {84}},\ \bibinfo
  {pages} {020508} (\bibinfo {year} {2011})}\BibitemShut {NoStop}%
\bibitem [{\citenamefont {Giovannetti}\ and\ \citenamefont
  {Capone}(2011)}]{giovannetti11}%
  \BibitemOpen
  \bibfield  {author} {\bibinfo {author} {\bibfnamefont {G.}~\bibnamefont
  {Giovannetti}}\ and\ \bibinfo {author} {\bibfnamefont {M.}~\bibnamefont
  {Capone}},\ }\href {\doibase 10.1103/PhysRevB.83.134508} {\bibfield
  {journal} {\bibinfo  {journal} {Phys.~Rev.~B}\ }\textbf {\bibinfo {volume}
  {83}},\ \bibinfo {pages} {134508} (\bibinfo {year} {2011})}\BibitemShut
  {NoStop}%
\bibitem [{\citenamefont {Casula}\ \emph {et~al.}(2011)\citenamefont {Casula},
  \citenamefont {Calandra}, \citenamefont {Profeta},\ and\ \citenamefont
  {Mauri}}]{casula11}%
  \BibitemOpen
  \bibfield  {author} {\bibinfo {author} {\bibfnamefont {M.}~\bibnamefont
  {Casula}}, \bibinfo {author} {\bibfnamefont {M.}~\bibnamefont {Calandra}},
  \bibinfo {author} {\bibfnamefont {G.}~\bibnamefont {Profeta}}, \ and\
  \bibinfo {author} {\bibfnamefont {F.}~\bibnamefont {Mauri}},\ }\href
  {\doibase 10.1103/PhysRevLett.107.137006} {\bibfield  {journal} {\bibinfo
  {journal} {Phys.~Rev.~Lett.}\ }\textbf {\bibinfo {volume} {107}},\ \bibinfo
  {pages} {137006} (\bibinfo {year} {2011})}\BibitemShut {NoStop}%
\bibitem [{\citenamefont {Kosugi}\ \emph {et~al.}(2011)\citenamefont {Kosugi},
  \citenamefont {Miyake}, \citenamefont {Ishibashi}, \citenamefont {Arita},\
  and\ \citenamefont {Aoki}}]{kosugi11}%
  \BibitemOpen
  \bibfield  {author} {\bibinfo {author} {\bibfnamefont {T.}~\bibnamefont
  {Kosugi}}, \bibinfo {author} {\bibfnamefont {T.}~\bibnamefont {Miyake}},
  \bibinfo {author} {\bibfnamefont {S.}~\bibnamefont {Ishibashi}}, \bibinfo
  {author} {\bibfnamefont {R.}~\bibnamefont {Arita}}, \ and\ \bibinfo {author}
  {\bibfnamefont {H.}~\bibnamefont {Aoki}},\ }\href {\doibase
  10.1103/PhysRevB.84.214506} {\bibfield  {journal} {\bibinfo  {journal}
  {Phys.~Rev.~B}\ }\textbf {\bibinfo {volume} {84}},\ \bibinfo {pages} {214506}
  (\bibinfo {year} {2011})}\BibitemShut {NoStop}%
\bibitem [{\citenamefont {de~Andres}\ \emph {et~al.}(2011)\citenamefont
  {de~Andres}, \citenamefont {Guijarro},\ and\ \citenamefont
  {Verg\'es}}]{Andres11}%
  \BibitemOpen
  \bibfield  {author} {\bibinfo {author} {\bibfnamefont {P.~L.}\ \bibnamefont
  {de~Andres}}, \bibinfo {author} {\bibfnamefont {A.}~\bibnamefont {Guijarro}},
  \ and\ \bibinfo {author} {\bibfnamefont {J.~A.}\ \bibnamefont {Verg\'es}},\
  }\href {\doibase 10.1103/PhysRevB.84.144501} {\bibfield  {journal} {\bibinfo
  {journal} {Phys.~Rev.~B}\ }\textbf {\bibinfo {volume} {84}},\ \bibinfo
  {pages} {144501} (\bibinfo {year} {2011})}\BibitemShut {NoStop}%
\bibitem [{\citenamefont {Kosugi}\ \emph {et~al.}(2009)\citenamefont {Kosugi},
  \citenamefont {Miyake}, \citenamefont {Ishibashi}, \citenamefont {Arita},\
  and\ \citenamefont {Aoki}}]{Picene_abinitio}%
  \BibitemOpen
  \bibfield  {author} {\bibinfo {author} {\bibfnamefont {T.}~\bibnamefont
  {Kosugi}}, \bibinfo {author} {\bibfnamefont {T.}~\bibnamefont {Miyake}},
  \bibinfo {author} {\bibfnamefont {S.}~\bibnamefont {Ishibashi}}, \bibinfo
  {author} {\bibfnamefont {R.}~\bibnamefont {Arita}}, \ and\ \bibinfo {author}
  {\bibfnamefont {H.}~\bibnamefont {Aoki}},\ }\href {\doibase
  10.1143/JPSJ.78.113704} {\bibfield  {journal} {\bibinfo  {journal}
  {J.~Phys.~Soc.~Jpn.}\ }\textbf {\bibinfo {volume} {78}},\ \bibinfo {pages}
  {113704} (\bibinfo {year} {2009})}\BibitemShut {NoStop}%
\bibitem [{\citenamefont {Huang}\ \emph {et~al.}(2012)\citenamefont {Huang},
  \citenamefont {Zhang},\ and\ \citenamefont {Lin}}]{huang12}%
  \BibitemOpen
  \bibfield  {author} {\bibinfo {author} {\bibfnamefont {Z.}~\bibnamefont
  {Huang}}, \bibinfo {author} {\bibfnamefont {C.}~\bibnamefont {Zhang}}, \ and\
  \bibinfo {author} {\bibfnamefont {H.-Q.}\ \bibnamefont {Lin}},\ }\href
  {\doibase 10.1038/srep00922} {\bibfield  {journal} {\bibinfo  {journal}
  {Sci.~Rep.}\ }\textbf {\bibinfo {volume} {2}},\ \bibinfo {pages} {992}
  (\bibinfo {year} {2012})}\BibitemShut {NoStop}%
\bibitem [{\citenamefont {Hebard}\ \emph {et~al.}(1991)\citenamefont {Hebard},
  \citenamefont {Rosseinsky}, \citenamefont {Haddon}, \citenamefont {Murphy},
  \citenamefont {Glarum}, \citenamefont {Palstra}, \citenamefont {Ramirez},\
  and\ \citenamefont {Kortan}}]{hebard91}%
  \BibitemOpen
  \bibfield  {author} {\bibinfo {author} {\bibfnamefont {A.~F.}\ \bibnamefont
  {Hebard}}, \bibinfo {author} {\bibfnamefont {M.~J.}\ \bibnamefont
  {Rosseinsky}}, \bibinfo {author} {\bibfnamefont {R.~C.}\ \bibnamefont
  {Haddon}}, \bibinfo {author} {\bibfnamefont {D.~W.}\ \bibnamefont {Murphy}},
  \bibinfo {author} {\bibfnamefont {S.~H.}\ \bibnamefont {Glarum}}, \bibinfo
  {author} {\bibfnamefont {T.~T.~M.}\ \bibnamefont {Palstra}}, \bibinfo
  {author} {\bibfnamefont {A.~P.}\ \bibnamefont {Ramirez}}, \ and\ \bibinfo
  {author} {\bibfnamefont {A.~R.}\ \bibnamefont {Kortan}},\ }\href {\doibase
  10.1038/350600a0} {\bibfield  {journal} {\bibinfo  {journal} {Nature}\
  }\textbf {\bibinfo {volume} {350}},\ \bibinfo {pages} {600} (\bibinfo {year}
  {1991})}\BibitemShut {NoStop}%
\bibitem [{\citenamefont {Capone}\ \emph {et~al.}(2002)\citenamefont {Capone},
  \citenamefont {Fabrizio}, \citenamefont {Castellani},\ and\ \citenamefont
  {Tosatti}}]{capone02}%
  \BibitemOpen
  \bibfield  {author} {\bibinfo {author} {\bibfnamefont {M.}~\bibnamefont
  {Capone}}, \bibinfo {author} {\bibfnamefont {M.}~\bibnamefont {Fabrizio}},
  \bibinfo {author} {\bibfnamefont {C.}~\bibnamefont {Castellani}}, \ and\
  \bibinfo {author} {\bibfnamefont {E.}~\bibnamefont {Tosatti}},\ }\href
  {\doibase 10.1126/science.1071122} {\bibfield  {journal} {\bibinfo  {journal}
  {Science}\ }\textbf {\bibinfo {volume} {296}},\ \bibinfo {pages} {2364}
  (\bibinfo {year} {2002})}\BibitemShut {NoStop}%
\bibitem [{\citenamefont {Ganin}\ \emph {et~al.}(2008)\citenamefont {Ganin},
  \citenamefont {Takabayashi}, \citenamefont {Khimyak}, \citenamefont
  {Margadonna}, \citenamefont {Tamai}, \citenamefont {Rosseinsky},\ and\
  \citenamefont {Prassides}}]{ganin08}%
  \BibitemOpen
  \bibfield  {author} {\bibinfo {author} {\bibfnamefont {A.~Y.}\ \bibnamefont
  {Ganin}}, \bibinfo {author} {\bibfnamefont {Y.}~\bibnamefont {Takabayashi}},
  \bibinfo {author} {\bibfnamefont {Y.~Z.}\ \bibnamefont {Khimyak}}, \bibinfo
  {author} {\bibfnamefont {S.}~\bibnamefont {Margadonna}}, \bibinfo {author}
  {\bibfnamefont {A.}~\bibnamefont {Tamai}}, \bibinfo {author} {\bibfnamefont
  {M.~J.}\ \bibnamefont {Rosseinsky}}, \ and\ \bibinfo {author} {\bibfnamefont
  {K.}~\bibnamefont {Prassides}},\ }\href {\doibase 10.1038/nmat2179}
  {\bibfield  {journal} {\bibinfo  {journal} {Nature Materials}\ }\textbf
  {\bibinfo {volume} {7}},\ \bibinfo {pages} {367} (\bibinfo {year}
  {2008})}\BibitemShut {NoStop}%
\bibitem [{\citenamefont {Ganin}\ \emph {et~al.}(2010)\citenamefont {Ganin},
  \citenamefont {Takabayashi}, \citenamefont {Jeglič}, \citenamefont {Arčon},
  \citenamefont {Potočnik}, \citenamefont {Baker}, \citenamefont {Ohishi},
  \citenamefont {McDonald}, \citenamefont {Tzirakis}, \citenamefont
  {Mc{L}ennan}, \citenamefont {Darling}, \citenamefont {Takata}, \citenamefont
  {Rosseinsky},\ and\ \citenamefont {Prassides}}]{ganin10}%
  \BibitemOpen
  \bibfield  {author} {\bibinfo {author} {\bibfnamefont {A.~Y.}\ \bibnamefont
  {Ganin}}, \bibinfo {author} {\bibfnamefont {Y.}~\bibnamefont {Takabayashi}},
  \bibinfo {author} {\bibfnamefont {P.}~\bibnamefont {Jeglič}}, \bibinfo
  {author} {\bibfnamefont {D.}~\bibnamefont {Arčon}}, \bibinfo {author}
  {\bibfnamefont {A.}~\bibnamefont {Potočnik}}, \bibinfo {author}
  {\bibfnamefont {P.~J.}\ \bibnamefont {Baker}}, \bibinfo {author}
  {\bibfnamefont {Y.}~\bibnamefont {Ohishi}}, \bibinfo {author} {\bibfnamefont
  {M.~T.}\ \bibnamefont {McDonald}}, \bibinfo {author} {\bibfnamefont {M.~D.}\
  \bibnamefont {Tzirakis}}, \bibinfo {author} {\bibfnamefont {A.}~\bibnamefont
  {Mc{L}ennan}}, \bibinfo {author} {\bibfnamefont {G.~R.}\ \bibnamefont
  {Darling}}, \bibinfo {author} {\bibfnamefont {M.}~\bibnamefont {Takata}},
  \bibinfo {author} {\bibfnamefont {M.~J.}\ \bibnamefont {Rosseinsky}}, \ and\
  \bibinfo {author} {\bibfnamefont {K.}~\bibnamefont {Prassides}},\ }\href
  {\doibase 10.1038/nature09120} {\bibfield  {journal} {\bibinfo  {journal}
  {Nature}\ }\textbf {\bibinfo {volume} {466}},\ \bibinfo {pages} {221}
  (\bibinfo {year} {2010})}\BibitemShut {NoStop}%
\bibitem [{\citenamefont {Capone}\ \emph {et~al.}(2009)\citenamefont {Capone},
  \citenamefont {Fabrizio}, \citenamefont {Castellani},\ and\ \citenamefont
  {Tosatti}}]{caponeRPM09}%
  \BibitemOpen
  \bibfield  {author} {\bibinfo {author} {\bibfnamefont {M.}~\bibnamefont
  {Capone}}, \bibinfo {author} {\bibfnamefont {M.}~\bibnamefont {Fabrizio}},
  \bibinfo {author} {\bibfnamefont {C.}~\bibnamefont {Castellani}}, \ and\
  \bibinfo {author} {\bibfnamefont {E.}~\bibnamefont {Tosatti}},\ }\href
  {\doibase 10.1103/RevModPhys.81.943} {\bibfield  {journal} {\bibinfo
  {journal} {Rev. Mod. Phys.}\ }\textbf {\bibinfo {volume} {81}},\ \bibinfo
  {pages} {943} (\bibinfo {year} {2009})}\BibitemShut {NoStop}%
\bibitem [{\citenamefont {et~al}(2009)}]{pwscf}%
  \BibitemOpen
  \bibfield  {author} {\bibinfo {author} {\bibfnamefont {P.~G.}\ \bibnamefont
  {et~al}},\ }\href {\doibase 10.1088/0953-8984/21/39/395502} {\bibfield
  {journal} {\bibinfo  {journal} {J.~Phys.~Condens.~Matter}\ }\textbf {\bibinfo
  {volume} {21}},\ \bibinfo {pages} {395502} (\bibinfo {year}
  {2009})}\BibitemShut {NoStop}%
\bibitem [{\citenamefont {Perdew}\ \emph {et~al.}(1997)\citenamefont {Perdew},
  \citenamefont {Burke},\ and\ \citenamefont {Ernzerhof}}]{PBE}%
  \BibitemOpen
  \bibfield  {author} {\bibinfo {author} {\bibfnamefont {J.~P.}\ \bibnamefont
  {Perdew}}, \bibinfo {author} {\bibfnamefont {K.}~\bibnamefont {Burke}}, \
  and\ \bibinfo {author} {\bibfnamefont {M.}~\bibnamefont {Ernzerhof}},\ }\href
  {\doibase 10.1103/PhysRevLett.78.1396} {\bibfield  {journal} {\bibinfo
  {journal} {Phys.~Rev.~Lett.}\ }\textbf {\bibinfo {volume} {78}},\ \bibinfo
  {pages} {1396} (\bibinfo {year} {1997})}\BibitemShut {NoStop}%
\bibitem [{\citenamefont {Vanderbilt}(1990)}]{PP-VDB}%
  \BibitemOpen
  \bibfield  {author} {\bibinfo {author} {\bibfnamefont {D.}~\bibnamefont
  {Vanderbilt}},\ }\href {\doibase 10.1103/PhysRevB.41.7892} {\bibfield
  {journal} {\bibinfo  {journal} {Phys. Rev. B}\ }\textbf {\bibinfo {volume}
  {41}},\ \bibinfo {pages} {7892} (\bibinfo {year} {1990})}\BibitemShut
  {NoStop}%
\bibitem [{\citenamefont {Trotter}(1963)}]{phen-crystal_siss}%
  \BibitemOpen
  \bibfield  {author} {\bibinfo {author} {\bibfnamefont {J.}~\bibnamefont
  {Trotter}},\ }\href {\doibase 10.1107/S0365110X63001626} {\bibfield
  {journal} {\bibinfo  {journal} {Acta Cryst.}\ }\textbf {\bibinfo {volume}
  {16}},\ \bibinfo {pages} {605} (\bibinfo {year} {1963})}\BibitemShut
  {NoStop}%
\bibitem [{\citenamefont {Dion}\ \emph {et~al.}(2004)\citenamefont {Dion},
  \citenamefont {Rydberg}, \citenamefont {Schr\"oder}, \citenamefont
  {Langreth},\ and\ \citenamefont {Lundqvist}}]{vdW-DF_SISS}%
  \BibitemOpen
  \bibfield  {author} {\bibinfo {author} {\bibfnamefont {M.}~\bibnamefont
  {Dion}}, \bibinfo {author} {\bibfnamefont {H.}~\bibnamefont {Rydberg}},
  \bibinfo {author} {\bibfnamefont {E.}~\bibnamefont {Schr\"oder}}, \bibinfo
  {author} {\bibfnamefont {D.~C.}\ \bibnamefont {Langreth}}, \ and\ \bibinfo
  {author} {\bibfnamefont {B.~I.}\ \bibnamefont {Lundqvist}},\ }\href {\doibase
  10.1103/PhysRevLett.92.246401} {\bibfield  {journal} {\bibinfo  {journal}
  {Phys. Rev. Lett.}\ }\textbf {\bibinfo {volume} {92}},\ \bibinfo {pages}
  {246401} (\bibinfo {year} {2004})}\BibitemShut {NoStop}%
\bibitem [{\citenamefont {Lee}\ \emph {et~al.}(2010)\citenamefont {Lee},
  \citenamefont {Murray}, \citenamefont {Kong}, \citenamefont {Lundqvist},\
  and\ \citenamefont {Langreth}}]{vdW-DF2}%
  \BibitemOpen
  \bibfield  {author} {\bibinfo {author} {\bibfnamefont {K.}~\bibnamefont
  {Lee}}, \bibinfo {author} {\bibfnamefont {E.~D.}\ \bibnamefont {Murray}},
  \bibinfo {author} {\bibfnamefont {L.}~\bibnamefont {Kong}}, \bibinfo {author}
  {\bibfnamefont {B.~I.}\ \bibnamefont {Lundqvist}}, \ and\ \bibinfo {author}
  {\bibfnamefont {D.~C.}\ \bibnamefont {Langreth}},\ }\href {\doibase
  10.1103/PhysRevB.82.081101} {\bibfield  {journal} {\bibinfo  {journal}
  {Phys.~Rev.~B}\ }\textbf {\bibinfo {volume} {82}},\ \bibinfo {pages} {081101}
  (\bibinfo {year} {2010})}\BibitemShut {NoStop}%
\bibitem [{\citenamefont {Grimme}(2004)}]{vdW-parameters}%
  \BibitemOpen
  \bibfield  {author} {\bibinfo {author} {\bibfnamefont {S.}~\bibnamefont
  {Grimme}},\ }\href {\doibase 10.1002/jcc.20078} {\bibfield  {journal}
  {\bibinfo  {journal} {J.~Comput.~Chem.}\ }\textbf {\bibinfo {volume} {25}},\
  \bibinfo {pages} {1463} (\bibinfo {year} {2004})}\BibitemShut {NoStop}%
\bibitem [{\citenamefont {Grimme}(2006)}]{vdW-Grimme_SISS}%
  \BibitemOpen
  \bibfield  {author} {\bibinfo {author} {\bibfnamefont {S.}~\bibnamefont
  {Grimme}},\ }\href {\doibase 10.1002/jcc.20495} {\bibfield  {journal}
  {\bibinfo  {journal} {J.~Comput.~Chem.}\ }\textbf {\bibinfo {volume} {27}},\
  \bibinfo {pages} {1787} (\bibinfo {year} {2006})}\BibitemShut {NoStop}%
\bibitem [{\citenamefont {Thonhauser}\ \emph {et~al.}(2007)\citenamefont
  {Thonhauser}, \citenamefont {Cooper}, \citenamefont {Li}, \citenamefont
  {Puzder}, \citenamefont {Hyldgaard},\ and\ \citenamefont
  {Langreth}}]{vdW-DF_SIS}%
  \BibitemOpen
  \bibfield  {author} {\bibinfo {author} {\bibfnamefont {T.}~\bibnamefont
  {Thonhauser}}, \bibinfo {author} {\bibfnamefont {V.~R.}\ \bibnamefont
  {Cooper}}, \bibinfo {author} {\bibfnamefont {S.}~\bibnamefont {Li}}, \bibinfo
  {author} {\bibfnamefont {A.}~\bibnamefont {Puzder}}, \bibinfo {author}
  {\bibfnamefont {P.}~\bibnamefont {Hyldgaard}}, \ and\ \bibinfo {author}
  {\bibfnamefont {D.~C.}\ \bibnamefont {Langreth}},\ }\href {\doibase
  10.1103/PhysRevB.76.125112} {\bibfield  {journal} {\bibinfo  {journal} {Phys.
  Rev. B}\ }\textbf {\bibinfo {volume} {76}},\ \bibinfo {pages} {125112}
  (\bibinfo {year} {2007})}\BibitemShut {NoStop}%
\bibitem [{\citenamefont {Rom\'an-P\'erez}\ and\ \citenamefont
  {Soler}(2009)}]{vdW-DF_SI}%
  \BibitemOpen
  \bibfield  {author} {\bibinfo {author} {\bibfnamefont {G.}~\bibnamefont
  {Rom\'an-P\'erez}}\ and\ \bibinfo {author} {\bibfnamefont {J.~M.}\
  \bibnamefont {Soler}},\ }\href {\doibase 10.1103/PhysRevLett.103.096102}
  {\bibfield  {journal} {\bibinfo  {journal} {Phys.~Rev.~Lett.}\ }\textbf
  {\bibinfo {volume} {103}},\ \bibinfo {pages} {096102} (\bibinfo {year}
  {2009})}\BibitemShut {NoStop}%
\bibitem [{\citenamefont {Sabatini}\ \emph {et~al.}(2012)\citenamefont
  {Sabatini}, \citenamefont {K\"uc\"ukbenli}, \citenamefont {B.Kolb},
  \citenamefont {Thonhauser},\ and\ \citenamefont
  {de~Gironcoli}}]{vdW-Stress-pwscf}%
  \BibitemOpen
  \bibfield  {author} {\bibinfo {author} {\bibfnamefont {R.}~\bibnamefont
  {Sabatini}}, \bibinfo {author} {\bibfnamefont {E.}~\bibnamefont
  {K\"uc\"ukbenli}}, \bibinfo {author} {\bibnamefont {B.Kolb}}, \bibinfo
  {author} {\bibfnamefont {T.}~\bibnamefont {Thonhauser}}, \ and\ \bibinfo
  {author} {\bibfnamefont {S.}~\bibnamefont {de~Gironcoli}},\ }\href {\doibase
  10.1088/0953-8984/24/42/424209} {\bibfield  {journal} {\bibinfo  {journal}
  {J.~Phys.~Condens.~Matter}\ }\textbf {\bibinfo {volume} {24}},\ \bibinfo
  {pages} {424209} (\bibinfo {year} {2012})}\BibitemShut {NoStop}%
\bibitem [{Note1()}]{Note1}%
  \BibitemOpen
  \bibinfo {note} {We thank A. Laio for this suggestion}\BibitemShut {NoStop}%
\bibitem [{Note2()}]{Note2}%
  \BibitemOpen
  \bibinfo {note} {We are grateful to Prof.~R.~Martonak for his help with these
  simulations}\BibitemShut {NoStop}%
\bibitem [{\citenamefont {Heine}(1960)}]{heine60}%
  \BibitemOpen
  \bibfield  {author} {\bibinfo {author} {\bibfnamefont {V.}~\bibnamefont
  {Heine}},\ }\href@noop {} {\emph {\bibinfo {title} {Group Theory in Quantum
  Mechanics}}}\ (\bibinfo  {publisher} {Pergamon Press, London},\ \bibinfo {year}
  {1960})\BibitemShut {NoStop}%
\bibitem [{\citenamefont {Bassani}\ and\ \citenamefont
  {Parravicin}(1975)}]{bassani75}%
  \BibitemOpen
  \bibfield  {author} {\bibinfo {author} {\bibfnamefont {F.}~\bibnamefont
  {Bassani}}\ and\ \bibinfo {author} {\bibfnamefont {G.~P.}\ \bibnamefont
  {Parravicin}},\ }\href@noop {} {\emph {\bibinfo {title} {Electron States and
  Optical Transitions in Solids}}}\ (\bibinfo  {publisher} {Pergamon Press, London},\
  \bibinfo {year} {1975})\BibitemShut {NoStop}%
\bibitem [{\citenamefont {Casula}\ \emph {et~al.}(2012)\citenamefont {Casula},
  \citenamefont {Calandra},\ and\ \citenamefont {Mauri}}]{casula12}%
  \BibitemOpen
  \bibfield  {author} {\bibinfo {author} {\bibfnamefont {M.}~\bibnamefont
  {Casula}}, \bibinfo {author} {\bibfnamefont {M.}~\bibnamefont {Calandra}}, \
  and\ \bibinfo {author} {\bibfnamefont {F.}~\bibnamefont {Mauri}},\ }\href
  {\doibase 10.1103/PhysRevB.86.075445} {\bibfield  {journal} {\bibinfo
  {journal} {Phys. Rev. B}\ }\textbf {\bibinfo {volume} {86}},\ \bibinfo
  {pages} {075445} (\bibinfo {year} {2012})}\BibitemShut {NoStop}%
\bibitem [{\citenamefont {Nomura}\ \emph {et~al.}(2012)\citenamefont {Nomura},
  \citenamefont {Nakamura},\ and\ \citenamefont {Arita}}]{nomura12}%
  \BibitemOpen
  \bibfield  {author} {\bibinfo {author} {\bibfnamefont {Y.}~\bibnamefont
  {Nomura}}, \bibinfo {author} {\bibfnamefont {K.}~\bibnamefont {Nakamura}}, \
  and\ \bibinfo {author} {\bibfnamefont {R.}~\bibnamefont {Arita}},\ }\href
  {\doibase 10.1103/PhysRevB.85.155452} {\bibfield  {journal} {\bibinfo
  {journal} {Phys. Rev. B}\ }\textbf {\bibinfo {volume} {85}},\ \bibinfo
  {pages} {155452} (\bibinfo {year} {2012})}\BibitemShut {NoStop}%
\bibitem [{\citenamefont {Mostofi}\ \emph {et~al.}(2008)\citenamefont
  {Mostofi}, \citenamefont {Yates}, \citenamefont {Lee}, \citenamefont
  {Souzab}, \citenamefont {Vanderbiltd},\ and\ \citenamefont
  {Marzari}}]{wannier90}%
  \BibitemOpen
  \bibfield  {author} {\bibinfo {author} {\bibfnamefont {A.~A.}\ \bibnamefont
  {Mostofi}}, \bibinfo {author} {\bibfnamefont {J.~R.}\ \bibnamefont {Yates}},
  \bibinfo {author} {\bibfnamefont {Y.-S.}\ \bibnamefont {Lee}}, \bibinfo
  {author} {\bibfnamefont {I.}~\bibnamefont {Souzab}}, \bibinfo {author}
  {\bibfnamefont {D.}~\bibnamefont {Vanderbiltd}}, \ and\ \bibinfo {author}
  {\bibfnamefont {N.}~\bibnamefont {Marzari}},\ }\href {\doibase
  10.1016/j.cpc.2007.11.016} {\bibfield  {journal} {\bibinfo
  {journal} { Comput. Phys. Commun.}\ } \textbf {\bibinfo {volume} {178}},\ \bibinfo
  {pages} {685} (\bibinfo {year} {2008})}\BibitemShut {NoStop}%
\bibitem [{\citenamefont {Shukla}\ \emph {et~al.}(2003)\citenamefont {Shukla},
  \citenamefont {Calandra}, \citenamefont {d'Astuto}, \citenamefont {Lazzeri},
  \citenamefont {Mauri}, \citenamefont {Bellin}, \citenamefont {Krisch},
  \citenamefont {Karpinski}, \citenamefont {Kazakov}, \citenamefont {Jun},
  \citenamefont {Daghero},\ and\ \citenamefont {Parlinski}}]{shukla03}%
  \BibitemOpen
  \bibfield  {author} {\bibinfo {author} {\bibfnamefont {A.}~\bibnamefont
  {Shukla}}, \bibinfo {author} {\bibfnamefont {M.}~\bibnamefont {Calandra}},
  \bibinfo {author} {\bibfnamefont {M.}~\bibnamefont {d'Astuto}}, \bibinfo
  {author} {\bibfnamefont {M.}~\bibnamefont {Lazzeri}}, \bibinfo {author}
  {\bibfnamefont {F.}~\bibnamefont {Mauri}}, \bibinfo {author} {\bibfnamefont
  {C.}~\bibnamefont {Bellin}}, \bibinfo {author} {\bibfnamefont
  {M.}~\bibnamefont {Krisch}}, \bibinfo {author} {\bibfnamefont
  {J.}~\bibnamefont {Karpinski}}, \bibinfo {author} {\bibfnamefont {S.~M.}\
  \bibnamefont {Kazakov}}, \bibinfo {author} {\bibfnamefont {J.}~\bibnamefont
  {Jun}}, \bibinfo {author} {\bibfnamefont {D.}~\bibnamefont {Daghero}}, \ and\
  \bibinfo {author} {\bibfnamefont {K.}~\bibnamefont {Parlinski}},\ }\href
  {\doibase 10.1103/PhysRevLett.90.095506} {\bibfield  {journal} {\bibinfo
  {journal} {Phys.~Rev.~Lett.}\ }\textbf {\bibinfo {volume} {90}},\ \bibinfo
  {pages} {095506} (\bibinfo {year} {2003})}\BibitemShut {NoStop}%
\bibitem [{\citenamefont {Fr\"ohlich}(1950)}]{Frol1_50}%
  \BibitemOpen
  \bibfield  {author} {\bibinfo {author} {\bibfnamefont {H.}~\bibnamefont
  {Fr\"ohlich}},\ }\href {\doibase 10.1103/PhysRev.79.845} {\bibfield
  {journal} {\bibinfo  {journal} {Phys. Rev.}\ }\textbf {\bibinfo {volume}
  {79}},\ \bibinfo {pages} {845} (\bibinfo {year} {1950})}\BibitemShut
  {NoStop}%
\bibitem [{\citenamefont {Fr\"ohlich}(1952)}]{Frol2_50}%
  \BibitemOpen
  \bibfield  {author} {\bibinfo {author} {\bibfnamefont {H.}~\bibnamefont
  {Fr\"ohlich}},\ }\href {\doibase 10.1098/rspa.1952.0212} {\bibfield
  {journal} {\bibinfo  {journal} {Proc.~Roy.~Soc.~London}\ }\textbf {\bibinfo
  {volume} {215}},\ \bibinfo {pages} {291} (\bibinfo {year}
  {1952})}\BibitemShut {NoStop}%
\end{thebibliography}
%

\end{document}